\documentclass[prd,preprintnumbers,twocolumn,amsmath,nofootinbib,amssymb]{revtex4}

\usepackage{graphicx,color,dcolumn,booktabs,bm}
\usepackage{longtable,lscape}
\usepackage{makecell}
\usepackage{txfonts}
\usepackage{overpic}
\usepackage{amssymb}
\usepackage{epstopdf}
\usepackage{indentfirst}
\usepackage{feynmf}   
\usepackage{slashed}  
\usepackage{cases}
\usepackage{color}
\usepackage{float}
\usepackage{multirow}
\usepackage{ulem}
\usepackage{enumerate}
\usepackage{graphicx,color,dcolumn,booktabs,bm}
\usepackage{epsfig,dsfont,amssymb,amsmath,amsfonts,amsbsy,mathrsfs}

\graphicspath{{Figures/}} %

\usepackage{hyperref}
\hypersetup{colorlinks,citecolor=blue,anchorcolor=red,menucolor=red, linkcolor=red,filecolor=red,runcolor=red,urlcolor=blue,frenchlinks=true}

\makeatletter
\@addtoreset{equation}{section}
\makeatother

\allowdisplaybreaks

\begin{document}

\title{Prediction of doubly-charm hadronic molecules with double strange quarks}

\author{Fu-Lai Wang$^{1,2,3,4}$}
\email{wangfulai@lzu.edu.cn}
\author{Xiang Liu$^{1,2,3,4}$\footnote{Corresponding author}}
\email{xiangliu@lzu.edu.cn}
\affiliation{$^1$School of Physical Science and Technology, Lanzhou University, Lanzhou 730000, China\\
$^2$Lanzhou Center for Theoretical Physics, Key Laboratory of Theoretical Physics of Gansu Province, Key Laboratory of Quantum Theory and Applications of MoE, Gansu Provincial Research Center for Basic Disciplines of Quantum Physics, Lanzhou University,
Lanzhou 730000, China\\
$^3$MoE Frontiers Science Center for Rare Isotopes, Lanzhou University, Lanzhou 730000, China\\
$^4$Research Center for Hadron and CSR Physics, Lanzhou University and Institute of Modern Physics of CAS, Lanzhou 730000, China}

\begin{abstract}
In this work, we investigate whether the $T$-doublet charmed-strange mesons and their antiparticles can form hidden-charm hidden-strangeness molecular tetraquarks by applying the one-boson-exchange model. We identify $D_{s1}\bar D_{s1}$ ($J^{PC}=0^{++},\,1^{+-},\,2^{++}$), $D_{s1}\bar D_{s2}^*$ ($J^{PC}=1^{+\pm},\,2^{+\pm},\,3^{+\pm}$), and $D_{s2}^*\bar D_{s2}^*$ ($J^{PC}=0^{++},\,1^{+-},\,2^{++},\,3^{+-},\,4^{++}$) as promising hidden-charm hidden-strangeness molecular tetraquark candidates. Notably, the $D_{s1}\bar D_{s2}^*$ state with $J^{PC}=2^{+-}$ possesses exotic spin-parity quantum numbers forbidden for conventional mesons, providing a clean experimental signature for exotic hadrons. Moreover, the $D_{s2}^*\bar D_{s2}^*$ state with $J^{PC}=4^{++}$ is a rare high-spin hadronic molecule. We then extend the same framework to discuss the binding properties of the $T_s T_s$ systems and construct the mass spectrum of corresponding doubly-charm doubly-strangeness molecular tetraquarks. The promising candidates are $D_{s1}D_{s1}$ ($J^P=2^+$), $D_{s1}D_{s2}^*$ ($J^P=3^+$), and $D_{s2}^*D_{s2}^*$ ($J^P=4^+$), all of which are absolutely flavor-exotic. We encourage experimental searches for these predicted hadronic molecules, which would be a crucial step toward establishing doubly-charm molecular tetraquarks with strangeness $S=0$ or $2$.
\end{abstract}
\maketitle

\section{Introduction}\label{sec1}

The study of exotic hadrons, the states that cannot be described as conventional quark-antiquark mesons or three-quark baryons, represents a highly active frontier in hadron physics. Over the past two decades, experimental advances have led to the observation of numerous exotic hadronic candidates \cite{Liu:2013waa,Hosaka:2016pey,Chen:2016qju,Richard:2016eis,Lebed:2016hpi,Liu:2019zoy,Brambilla:2019esw,Chen:2022asf,Olsen:2017bmm,Guo:2017jvc,Meng:2022ozq,Liu:2024uxn,Wang:2025sic,Wang:2025dur,Bai:2026atm}. Unraveling the internal structures of these exotic hadrons goes beyond mere cataloging: it not only helps construct hadron spectroscopy but also deepens our understanding of non-perturbative strong interactions.

Among the observed exotic hadronic candidates, the charmonium-like $XYZ$ states form the largest group \cite{Brambilla:2019esw}. A notable feature of many $XYZ$ states is that their masses lie remarkably close to thresholds of two charmed mesons, a property that has fueled extensive studies of hidden-charm molecular tetraquarks \cite{Liu:2013waa,Hosaka:2016pey,Chen:2016qju,Richard:2016eis,Lebed:2016hpi,Liu:2019zoy,Brambilla:2019esw,Chen:2022asf,Olsen:2017bmm,Guo:2017jvc,Meng:2022ozq,Liu:2024uxn,Wang:2025sic,Wang:2025dur,Bai:2026atm}. Within this context, several hidden-charm hidden-strangeness molecular tetraquark candidates have been discussed. For instance, the $X(3960)$ \cite{LHCb:2022aki} and $Y(4140)$ \cite{Aaltonen:2009tz} lie near the $D_s\bar D_s$ and $D_s^*\bar D_s^*$ thresholds, respectively, and have been extensively studied as hadronic molecules formed by the $S$-wave charmed-strange mesons and their antiparticles \cite{Ji:2022uie,Bayar:2022dqa,Ding:2023yuo,Qi:2023gwb,Peng:2023lfw,Agaev:2023gti,Chen:2023eix,Mutuk:2022ckn,Xin:2022bzt,Zhang:2009st,Wang:2014gwa,Chen:2015fdn,Mahajan:2009pj,Ding:2009vd,Liu:2009ei,Branz:2009yt,Albuquerque:2009ak,Molina:2009ct,Liu:2010rnq}. Other states such as $X(4274)$ \cite{Aaltonen:2009tz,Aaij:2016iza}, $Y(4626)$ \cite{Belle:2019qoi}, and $X(4630)$ \cite{LHCb:2021uow} appear near thresholds involving one $S$-wave and one $P$-wave charmed-strange mesons, and their molecular interpretations have also been investigated \cite{Finazzo:2011he,Liu:2010hf,He:2013oma,He:2016pfa,Peng:2022nrj,Ke:2020eba,He:2019csk,Yang:2021sue,Wang:2021ghk}. These studies naturally raise a further question: can the $T$-doublet charmed-strange mesons\footnote{In this work, the $T$-doublet charmed-strange mesons $D_{s1}(2536)$ and $D_{s2}^*(2573)$ are denoted simply as $D_{s1}$ and $D_{s2}^*$ \cite{ParticleDataGroup:2024cfk}.} and their antiparticles also form hidden-charm hidden-strangeness molecular tetraquarks (denoted as $T_s \bar T_s$)? Addressing this question is the central goal of the present work.

Compared with molecular tetraquarks built from the $S$-wave charmed-strange mesons and their antiparticles, those composed of the $T$-doublet charmed-strange mesons and their antiparticles offer three distinctive advantages: (i) Exotic spin-parity quantum numbers. The combination of the $T$-doublet charmed-strange mesons and their antiparticles can yield total angular momentum and parity assignments, such as $J^{PC}=2^{+-}$, that are forbidden for conventional $c\bar c$ mesons \cite{Chen:2016qju}. Such exotic quantum numbers provide a clean signature for the exotic nature of a hadron, directly distinguishing them from conventional hadrons. In contrast, molecular tetraquarks formed from the $S$-wave charmed-strange mesons and their antiparticles only produce non-exotic quantum numbers, which could also arise from conventional $c\bar c$ mesons. This poses a dilemma for identifying exotic hadrons from the charmonium-like $XYZ$ states \cite{Liu:2013waa,Hosaka:2016pey,Chen:2016qju,Richard:2016eis,Lebed:2016hpi,Liu:2019zoy,Brambilla:2019esw,Chen:2022asf,Olsen:2017bmm,Guo:2017jvc,Meng:2022ozq,Liu:2024uxn,Wang:2025sic,Wang:2025dur,Bai:2026atm}. Thus, hidden-charm hidden-strangeness molecular tetraquarks composed of the $T$-doublet charmed-strange mesons and their antiparticles open a unique window for identifying genuine hadronic molecules via quantum numbers. (ii) Larger reduced mass favors binding. Within the one-boson-exchange (OBE) model, the interaction of hidden-charm hidden-strangeness tetraquark system is mediated by $f_0(980)$\footnote{For simplicity, the scalar meson $f_0(980)$ is denoted as $f_0$ throughout this work \cite{ParticleDataGroup:2024cfk}.}, $\eta$, and $\phi$ exchanges \cite{Liu:2013waa}. The $T$-doublet charmed-strange mesons $D_{s1}$ and $D_{s2}^*$ are significantly heavier than their $S$-wave partners, leading to a larger reduced mass. A larger reduced mass reduces the kinetic energy, making it easier for an attractive potential to bind the system \cite{Li:2014gra,Karliner:2015ina}. (iii) High spin up to $J=4$. The $T$-doublet charmed-strange mesons carry spins $1$ and $2$, allowing total angular momenta of systems composed of the $T$-doublet charmed-strange mesons and their antiparticles up to $J=4$ in the $S$-wave configurations. Currently, the highest spin observed in the $XYZ$ family is $J=2$ (e.g., $X(4150)$ \cite{LHCb:2021uow}), and high-spin hadronic molecules remain experimentally absent \cite{ParticleDataGroup:2024cfk}. The prediction of spin-3 and spin-4 molecular tetraquarks would therefore open a new window into the spectroscopy of high-spin exotic hadrons.

Beyond hidden-charm hidden-strangeness tetraquark systems, the discovery of $T_{cc}(3875)^+$ by LHCb in 2021 \cite{LHCb:2021vvq,LHCb:2021auc} provides a compelling example of a doubly-charm molecular tetraquark, interpreted as a $DD^*$ hadronic molecule \cite{Manohar:1992nd,Ericson:1993wy,Tornqvist:1993ng,Janc:2004qn,Ding:2009vj,Molina:2010tx,Ding:2020dio,Li:2012ss,Xu:2017tsr,Liu:2019stu,Ohkoda:2012hv,Tang:2019nwv}. This finding naturally raises an analogous question: can two $T$-doublet charmed-strange mesons form doubly-charm doubly-strangeness molecular $T_s T_s$ tetraquarks? Such $T_s T_s$ systems would carry open charm $C=+2$ and open strangeness $S=+2$, with a valence quark content $cc\bar s\bar s$ that is absolutely flavor-exotic, a configuration irreducible to conventional hadrons. Therefore, we extend our analysis to the $T_s T_s$ systems and construct the mass spectrum of these doubly-charm doubly-strangeness molecular tetraquarks.

In this work, we systematically investigate the mass spectrum of doubly-charm molecular tetraquarks with strangeness $S=0$ or $2$, formed from the $T$-doublet charmed-strange mesons. Two types of configurations are considered: hidden-charm hidden-strangeness $T_s \bar T_s$ tetraquarks and doubly-charm doubly-strangeness $T_s T_s$ tetraquarks. We employ the OBE model to derive the interaction potentials of the $T_s \bar T_s$ systems. These effective potentials are then inserted into the coupled-channel Schr\"odinger equation to search for the loosely bound state solutions, and the mass spectrum of the hidden-charm hidden-strangeness molecular $T_s \bar T_s$ tetraquarks is constructed accordingly. The same procedure is then applied to doubly-charm doubly-strangeness $T_s T_s$ tetraquark systems to determine their binding properties and mass spectrum.

The remainder of this paper is organized as follows. In Sec.~\ref{sec2}, we derive the OBE interaction potentials of the $T_s\bar T_s$ systems and solve the coupled-channel Schr\"odinger equation to construct the mass spectrum of these hidden-charm hidden-strangeness molecular tetraquarks. In Sec.~\ref{sec3}, we extend the analysis to doubly-charm doubly-strangeness $T_s T_s$ tetraquark systems and construct their mass spectrum. Finally, a brief discussion and conclusion is given in Sec.~\ref{sec4}.

\section{Hidden-charm hidden-strangeness molecular $T_s \bar T_s$ tetraquarks}\label{sec2}

In this section, we construct the mass spectrum of hidden-charm hidden-strangeness molecular tetraquarks formed from the $T$-doublet charmed-strange mesons and their antiparticles. The analysis proceeds in two stages: (i) We employ the OBE model, which has proven successful in describing hadronic molecules such as the deuteron, $P_c$, and $T_{cc}$ \cite{Chen:2016qju,Liu:2019zoy}, to derive the interaction potentials of the $T_s \bar T_s$ systems. This approach systematically incorporates interactions mediated by the scalar meson $f_0$, the pseudoscalar meson $\eta$, and the vector meson $\phi$. (ii) The OBE effective potentials are projected onto the partial-wave basis comprising the $S$ and $D$ waves and are then inserted into the coupled-channel Schr\"odinger equation. For each set of quantum numbers $J^{PC}$, we extract the binding energies, masses, root-mean-square (RMS) radii, and partial-wave compositions. From the obtained bound state solutions, we construct the mass spectrum of hidden-charm hidden-strangeness molecular $T_s \bar T_s$ tetraquarks.

\subsection{Effective potentials of the $T_s \bar T_s$ systems}

\subsubsection{Effective Lagrangians}

To derive the OBE effective potentials of the $T_s \bar T_s$ systems, we first construct the effective Lagrangians describing the couplings of the $T$-doublet charmed-strange mesons (and their antiparticles) to the light mesons $f_0$, $\eta$, and $\phi$. The $T$-doublet charmed-strange mesons consist of the $D_{s1}$ and $D_{s2}^*$ \cite{ParticleDataGroup:2024cfk}, which share the same light-quark angular momentum $j_q=3/2$. Within the heavy-quark effective theory \cite{Wise:1992hn}, the spin partners are combined into a superfield. For the $T$-doublet charmed-strange mesons and their antiparticles, we have \cite{Ding:2008gr}
\begin{eqnarray*}
T^{(c)\mu}_a&=&\frac{1+{v}\!\!\!\slash}{2}\left[D^{*\mu\nu}_{2a}\gamma_{\nu}-\sqrt{\frac{3}{2}}D_{1a\nu}\gamma_5\left(g^{\mu\nu}
-\frac{\gamma^{\nu}(\gamma^{\mu}-v^{\mu})}{3}\right)\right],\\
T^{(\bar c)\mu}_a&=&\left[\bar D^{*\mu\nu}_{2a}\gamma_{\nu}-\sqrt{\frac{3}{2}}\bar D_{1a\nu}\gamma_5\left(g^{\mu\nu}-\frac{(\gamma^{\mu}-v^{\mu})\gamma^{\nu}}{3}
\right)\right]\frac{1-\slash \!\!\!v}{2}.
\end{eqnarray*}
The corresponding conjugate fields are $\overline{T}^{(c)\mu} = \gamma^0 T^{(c)\mu\dagger}\gamma^0$ and similarly for the antiparticle. The fields contain the physical states: $D_1=(D_1^0,\,D_1^+,\,D_{s1}^+)$, $D_{2}^{*}=(D_2^{*0},\,D_2^{*+},\,D_{s2}^{*+})$, $\bar D_1=(\bar D_1^0,\,D_1^-,\,D_{s1}^-)$, and $\bar D_{2}^{*}=(\bar D_2^{*0},\,D_2^{*-},\,D_{s2}^{*-})$. The normalization relations of these discussed charmed-strange mesons are $\langle 0|D_{s1}^{\mu}|c\bar{s}(1^+)\rangle=\epsilon^{\mu}\sqrt{m_{D_{s1}}}$, $\langle 0|D_{s2}^{*\mu\nu}|c\bar{s}(2^+)\rangle=\xi^{\mu\nu}\sqrt{m_{D_{s2}^{\ast}}}$, and analogously for the antiparticles \cite{Ding:2008gr}. In the static limit, $v^\mu=(1,\,\bm{0})$, while the polarization vector $\epsilon^\mu_m$ ($m=0,\,\pm1$) and the polarization tensor $\zeta^{\mu\nu}_m$ ($m=0,\,\pm1,\,\pm2$) are given by \cite{Cheng:2010yd}
\begin{eqnarray*}
\epsilon_{0}^{\mu}&=&\left(0,0,0,-1\right),\nonumber\\
\epsilon_{\pm}^{\mu}&=&\frac{1}{\sqrt{2}}\left(0,\,\pm1,\,i,\,0\right),\nonumber\\
\zeta^{\mu\nu}_{m}&=&\sum_{m_1,m_2}C^{2, m}_{1m_1,1m_2}\epsilon^{\mu}_{m_1}\epsilon^{\nu}_{m_2}.
\end{eqnarray*}

Within the framework of the heavy quark symmetry, the chiral symmetry, and the hidden gauge symmetry \cite{Casalbuoni:1992gi,Casalbuoni:1996pg,Yan:1992gz,Harada:2003jx,Bando:1987br}, we construct the effective Lagrangians that describe the couplings of the $T$-doublet charmed-strange mesons (and their antiparticles) to the light mesons as follows \cite{Ding:2008gr}.
\paragraph{{\rm For the light scalar mesons:}}
\begin{eqnarray}
\mathcal{L}_{TTf_0}=g''_{f_0}\langle T^{(c)\mu}_af_0\overline{T}^{\,({c})}_{a\mu}\rangle
+g''_{f_0}\langle \overline{T}^{(\bar c)\mu}_af_0 T^{\,({\bar c})}_{a\mu}\rangle.
\end{eqnarray}
\paragraph{{\rm For the light pseudoscalar mesons:}}
\begin{eqnarray}
\mathcal{L}_{TT\mathbb{P}}=ig''\langle T^{(c)\mu}_{b}{\cal A}\!\!\!\slash_{ba}\gamma_5\overline{T}^{\,(c)}_{a\mu}\rangle+ig''\langle \overline{T}^{(\bar c)\mu}_{b}{\cal A}\!\!\!\slash_{ba}\gamma_5{T}^{\,(\bar c)}_{a\mu}\rangle,
\end{eqnarray}
where $\mathcal{A}_\mu = \frac{1}{2}(\xi^\dagger\partial_\mu\xi - \xi\partial_\mu\xi^\dagger)$ and $\xi = e^{i\mathbb{P}/f_\pi}$. The matrix $\mathbb{P}$ is
\begin{eqnarray*}
{\mathbb{P}} &=& {\left(\begin{array}{ccc}
       \frac{\pi^0}{\sqrt{2}}+\frac{\eta}{\sqrt{6}} &\pi^+ &K^+\\
       \pi^-       &-\frac{\pi^0}{\sqrt{2}}+\frac{\eta}{\sqrt{6}} &K^0\\
       K^-         &\bar K^0   &-\sqrt{\frac{2}{3}} \eta     \end{array}\right)}.
\end{eqnarray*}
\paragraph{{\rm For the light vector mesons:}}
\begin{eqnarray}
\mathcal{L}_{TT\mathbb{V}}&=&i\beta^{\prime\prime}\langle T^{(c)\lambda}_bv^{\mu}({\cal V}_{\mu}-\rho_{\mu})_{ba}\overline{T}^{\,(c)}_{a\lambda}\rangle\nonumber\\
&&+i\lambda^{\prime\prime}\langle T^{(c)\lambda}_b\sigma^{\mu\nu}F_{\mu\nu}(\rho)_{ba}\overline{T}^{\,(c)}_{a\lambda}\rangle\nonumber\\
&&-i\beta^{\prime\prime}\langle \overline{T}^{(\bar c)}_{b\lambda}v^{\mu}({\cal V}_{\mu}-\rho_{\mu})_{ba}T^{\,(\bar c)\lambda}_{a}\rangle\nonumber\\
    &&+i\lambda^{\prime\prime}\langle \overline{T}^{(\bar c)}_{b\lambda}\sigma^{\mu\nu}F_{\mu\nu}(\rho)_{ba}T^{\,(\bar c)\lambda}_{a}\rangle,
\end{eqnarray}
where $\mathcal{V}_\mu = \frac12(\xi^\dagger\partial_\mu\xi + \xi\partial_\mu\xi^\dagger)$, $F_{\mu\nu}(\rho)=\partial_\mu\rho_\nu-\partial_\nu\rho_\mu+[\rho_\mu,\rho_\nu]$, and $\rho_\mu = i g_V \mathbb{V}_\mu/\sqrt{2}$. The matrix $\mathbb{V}_\mu$ is
\begin{eqnarray*}
{\mathbb{V}}_{\mu} &=& {\left(\begin{array}{ccc}
       \frac{\rho^0}{\sqrt{2}}+\frac{\omega}{\sqrt{2}} &\rho^+ &K^{*+}\\
       \rho^-       &-\frac{\rho^0}{\sqrt{2}}+\frac{\omega}{\sqrt{2}} &K^{*0}\\
       K^{*-}         &\bar K^{*0}   & \phi     \end{array}\right)}_{\mu}.
\end{eqnarray*}

Expanding the effective Lagrangians above yields the following explicit expressions:
\begin{eqnarray}
\mathcal{L}_{D_1 D_1f_0} &=& -2g_{f_0}^{\prime\prime}D_{1a\mu}D^{\mu\dagger}_{1a} f_0 ,\\
\mathcal {L}_{D_1 D_1\mathbb{P}}&=&-\frac{5ik}{3f_\pi}~\epsilon^{\mu\nu\rho\tau}v_\tau D_{1b\nu}D^{\dagger}_{1a\mu} \partial_\rho\mathbb{P}_{ba},\\
\mathcal {L}_{D_1 D_1\mathbb{V}} &=& -\sqrt{2}\beta^{\prime \prime}g_{V}\left(v\cdot\mathbb{V}_{ba}\right) D_{1b\mu}D^{\mu\dagger}_{1a}\nonumber\\
&&+\frac{5\sqrt{2}i\lambda^{\prime\prime} g_{V}}{3}\left(D^{\mu}_{1b}D^{\nu\dagger}_{1a}-D^{\nu}_{1b}D^{\mu\dagger}_{1a}\right)\partial_\mu \mathbb{V}_{ba\nu},\\
\mathcal{L}_{D^{\ast}_2D^{\ast}_2f_0} &=& 2g_{f_0}^{\prime\prime}D^{*\mu\nu}_{2a}D^{*\dagger}_{2a\mu\nu}f_0 ,\\
\mathcal {L}_{D^{\ast}_2D^{\ast}_2\mathbb{P}}&=&\frac{2ik}{f_\pi}~\epsilon^{\mu\nu\rho\tau}v_\nu D^{*}_{2b\alpha\tau}D^{*\alpha\dagger}_{2a\rho}\partial_\mu\mathbb{P}_{ba},\\
\mathcal {L}_{D^{\ast}_2D^{\ast}_2\mathbb{V}} &=& \sqrt{2}\beta^{\prime \prime}g_{V}\left(v\cdot\mathbb{V}_{ba}\right) D_{2b}^{*\lambda\nu}  D^{*\dagger}_{2a{\lambda\nu}}+2\sqrt{2}i\lambda^{\prime\prime} g_{V}\nonumber\\
&&\times\left(D^{*\lambda\nu}_{2b} D^{*\mu\dagger}_{2a\lambda}-D^{*\mu}_{2b\lambda}D^{*\lambda\nu\dagger}_{2a}\right)\partial_\mu \mathbb{V}_{ba\nu},\\
\mathcal {L}_{D_1D^{\ast}_2\mathbb{P}}&=&-\sqrt{\frac{2}{3}}\frac{k}{f_\pi}\left(D^{*\mu\lambda}_{2b}D^{\dagger}_{1a\mu}
+D_{1b\mu}D^{*\mu\lambda\dagger}_{2a}\right)\partial_\lambda\mathbb{P}_{ba},\\
\mathcal {L}_{D_1D^{\ast}_2\mathbb{V}} &=& \frac{i\beta^{\prime \prime}g_{V}}{\sqrt{3}}\epsilon^{\lambda\alpha\rho\tau}v_{\rho}\left(v\cdot\mathbb{V}_{ba}\right)\left(D^{*}_{2b\lambda\tau}D^{\dagger}_{1a\alpha}-D_{1b\alpha}
D^{*\dagger}_{2a\lambda\tau}\right)\nonumber\\
&&+\frac{2\lambda^{\prime\prime} g_{V}}{\sqrt{3}}\left[3\epsilon^{\mu\lambda\nu\tau}v_\lambda\left(D^{*}_{2b\alpha\tau}D^{\alpha\dagger}_{1a}
+D^{\alpha}_{1b}D^{*\dagger}_{2a\alpha\tau}\right)\partial_\mu \mathbb{V}_{ba\nu}\right.\nonumber\\
&&+2\epsilon^{\lambda\alpha\rho\nu}v_\rho\left(D^{*\mu}_{2b\lambda}D^{\dagger}_{1a\alpha}+D_{1b\alpha}D^{*\mu\dagger}_{2a\lambda}\right)\nonumber\\
&&\left.\times\left(\partial_\mu \mathbb{V}_{ba\nu}-\partial_\nu \mathbb{V}_{ba\mu}\right)\right],\\
\mathcal{L}_{\bar D_1 \bar D_1f_0}&=&-2g_{f_0}^{\prime\prime}\bar{D}_{1a\mu}\bar{D}^{\mu\dagger}_{1a}f_0,\\
\mathcal {L}_{\bar D_1 \bar D_1\mathbb{P}} &=&-\frac{5ik}{3f_\pi}\varepsilon^{\mu\nu\rho\tau}v_\nu\bar{D}^{\dag}_{1a\rho}
\bar{D}_{1b\tau}\partial_\mu\mathbb{P}_{ba},\\
\mathcal {L}_{\bar D_1 \bar D_1\mathbb{V}} &=&\sqrt{2}\beta^{\prime\prime}g_{V}\left(v\cdot\mathbb{V}_{ba}\right)\bar{D}_{1b\mu}
\bar{D}^{\mu\dagger}_{1a}\nonumber\\
    &&+\frac{5\sqrt{2}i\lambda^{\prime\prime}g_{V}}{3}\left(\bar{D}^{\nu}_{1b}
    \bar{D}^{\mu\dagger}_{1a}
    -\bar{D}^{\nu\dagger}_{1a}\bar{D}^{\mu}_{1b}\right)\partial_\mu\mathbb{V}_{ba\nu},\\
\mathcal{L}_{\bar D^{\ast}_2 \bar D^{\ast}_2f_0}&=&2g_{f_0}^{\prime\prime}\bar{D}^{*\dagger}_{2a\mu\nu}\bar{D}^{*\mu\nu}_{2a}f_0,\\
\mathcal {L}_{\bar D^{\ast}_2 \bar D^{\ast}_2\mathbb{P}} &=&\frac{2ik}{f_\pi}\varepsilon^{\mu\nu\rho\tau}v_\nu\bar{D}^{*\alpha\dag}_{2a\rho}
\bar{D}^{*}_{2b\alpha\tau}\partial_\mu\mathbb{P}_{ba},\\
\mathcal {L}_{\bar D^{\ast}_2 \bar D^{\ast}_2\mathbb{V}} &=&-\sqrt{2}\beta^{\prime \prime}g_{V}\left(v\cdot\mathbb{V}_{ba}\right) \bar{D}_{2b}^{*\lambda\nu} \bar{D}^{*\dagger}_{2a{\lambda\nu}}\nonumber\\
    &&+2\sqrt{2}i\lambda^{\prime\prime} g_{V}\left(\bar{D}^{*\lambda\nu\dagger}_{2a}
    \bar{D}^{*\mu}_{2b\lambda}-\bar{D}^{*\lambda\nu}_{2b} \bar{D}^{*\mu\dagger}_{2a\lambda}\right)\partial_\mu \mathbb{V}_{ba\nu},\\
\mathcal {L}_{\bar D_1 \bar D^{\ast}_2\mathbb{P}} &=&\sqrt{\frac{2}{3}}\frac{k}{f_\pi}\left(\bar{D}^{\dagger}_{1a\mu}
    \bar{D}^{*\mu\lambda}_{2b}+\bar{D}_{1b\mu}   \bar{D}^{*\mu\lambda\dagger}_{2a}\right)\partial_\lambda\mathbb{P}_{ba},\\
\mathcal {L}_{\bar D_1 \bar D^{\ast}_2\mathbb{V}} &=&\frac{i\beta^{\prime\prime}g_{V}}{\sqrt{3}}\varepsilon^{\lambda\alpha\rho\tau}v_{\rho}
    \left(v\cdot\mathbb{V}_{ba}\right)\left(\bar{D}^{\dagger}_{1a\alpha} \bar{D}^{*}_{2b\lambda\tau}-\bar{D}_{1b\alpha}
    \bar{D}^{\dagger*}_{2a\lambda\tau}\right)\nonumber\\
    &&+\frac{2\lambda^{\prime\prime}g_{V}}{\sqrt{3}}   \left[3\varepsilon^{\mu\lambda\nu\tau}v_\lambda\left(\bar{D}^{\alpha\dagger}_{1a}
    \bar{D}^{*}_{2b\alpha\tau}+\bar{D}^{\alpha}_{1b} \bar{D}^{*\dagger}_{2a\alpha\tau}\right)\partial_\mu\mathbb{V}_{ba\nu}\right.\nonumber\\   &&\left.+2\varepsilon^{\lambda\alpha\rho\nu}v_\rho   \left(\bar{D}^{\dagger}_{1a\alpha}\bar{D}^{*\mu}_{2b\lambda}   +\bar{D}_{1b\alpha}\bar{D}^{\dagger\mu*}_{2a\lambda}\right)\right.\nonumber\\
    &&\left.\times\left(\partial_\mu \mathbb{V}_{ba\nu}-\partial_\nu \mathbb{V}_{ba\mu}\right)\right].
\end{eqnarray}

In the above effective Lagrangians, there exists a series of coupling constants. For the $T$-doublet charmed-strange mesons, direct experimental determinations of these couplings are currently unavailable \cite{ParticleDataGroup:2024cfk}. Fortunately, abundant experimental data exist for the nucleon-nucleon interactions \cite{Machleidt:1987hj,Machleidt:2000ge,Cao:2010km}. Therefore, in this work we employ the quark model~\cite{Riska:2000gd}, a common approach for determining such couplings, to relate the coupling constants of the $T$-doublet charmed-strange mesons to those of the nucleon-nucleon interaction vertices. The resulting numerical values are $g^{\prime\prime}_{f_0}=2.82$,  $g''=0.78$, $f_{\pi}=0.132~\rm{GeV}$, $\beta^{\prime\prime}g_V=6.50$, and $\lambda^{\prime\prime}g_V=7.38~\rm{GeV}^{-1}$. These values have been widely used in previous studies of the mass spectrum of hadronic molecular states related to the $T$-doublet charmed-strange mesons \cite{Wang:2019aoc,Cui:2026hfm,Qi:2021iyv,Wang:2020lua}. The meson masses are taken from the Particle Data Group \cite{ParticleDataGroup:2024cfk}: $m_{D_{s1}}=2531.21~\rm{MeV}$, $m_{D_{s2}^{*}}=2569.10~\rm{MeV}$, $m_{f_0}=990.00~\rm{MeV}$, $m_{\eta}=547.86~\rm{MeV}$, and $m_{\phi}=1019.46~\rm{MeV}$.

\subsubsection{Flavor and spin-orbital wave functions}

The flavor part of the wave function is determined by the quantum numbers of the constituents. Both $T$-doublet charmed-strange mesons are isoscalars with strangeness $S=-1$, while their antiparticles carry strangeness $S=1$. Consequently, the $T_s \bar T_s$ systems have total strangeness $S=0$ and total isospin $I=0$. For the $D_{s1}\bar D_{s1}$, $D_{s1}\bar D_{s2}^*$, and $D_{s2}^*\bar D_{s2}^*$ systems, the corresponding flavor wave functions $|I,I_3\rangle = |0,0\rangle$ are given by:
\begin{center}
\renewcommand{\arraystretch}{1.50}
\begin{tabular*}{80mm}{@{\extracolsep{\fill}}cccc}
\toprule[1.00pt]
~~Systems~~& $D_{s1} \bar D_{s1}$ & $D_{s1} \bar D_{s2}^{\ast}$ & $D_{s2}^{\ast}\bar D_{s2}^{\ast}$~~\\\hline
$|0,\,0\rangle$&$D_{s1}^+D_{s1}^-$&$D_{s1}^+ D_{s2}^{\ast-}+cD_{s2}^{\ast+}D_{s1}^-$&$D_{s2}^{\ast+}D_{s2}^{\ast-}$\\
\bottomrule[1.00pt]
\end{tabular*}.
\end{center}
For the $D_{s1}\bar D_{s2}^*$ system, the charge-conjugation transformation exchanges two $T$-doublet charmed-strange mesons \cite{Ding:2008gr}, yielding $C=-c(-1)^{J-3}$ with $c=\pm 1$ \cite{Liu:2007bf,Liu:2008xz,Liu:2008fh,Liu:2008tn,Sun:2012sy,Wang:2020dya,Li:2015exa,Li:2013bca,Hu:2010fg,Shen:2010ky,Dong:2021juy,Liu:2013rxa,Chen:2015add,Wang:2021aql}, where $C$ and $J$ denote the charge-conjugation parity and the total angular momentum quantum number of the $D_{s1}\bar D_{s2}^*$ system, respectively. This relation will be employed later to assign the $C$-parity of the $D_{s1} \bar D_{s2}^{\ast}$ system.

The spin-orbital part of the wave function is built by coupling the intrinsic spins of two $T$-doublet charmed-strange mesons with the relative orbital angular momentum to total angular momentum, and the form of the spin-orbital wave functions $|{}^{2S+1}L_{J}\rangle$ for the $D_{s1} \bar D_{s1}$, $D_{s1} \bar D_{s2}^{\ast}$, and $D_{s2}^{\ast} \bar D_{s2}^{\ast}$ systems can be expressed as
\begin{center}
\renewcommand{\arraystretch}{1.50}
\begin{tabular*}{80mm}{@{\extracolsep{\fill}}cc}
\toprule[1.00pt]
~~Systems~~&$|{}^{2S+1}L_{J}\rangle$ \\\hline
$D_{s1} \bar D_{s1}$&$\sum_{m,m^{\prime},m_S,m_L}C^{S,m_S}_{1m,1m^{\prime}}C^{J,M}_{Sm_S,Lm_L}\epsilon_{m}^\mu\epsilon_{m^{\prime}}^\nu|Y_{L\,m_L}\rangle$~~\nonumber\\
$D_{s1} \bar D_{s2}^{\ast}$&$\sum_{m,m^{\prime},m_S,m_L}C^{S,m_S}_{1m,2m^{\prime}}C^{J,M}_{Sm_S,Lm_L}\epsilon_{m}^\lambda
\zeta_{m^{\prime}}^{\mu\nu}|Y_{L\,m_L}\rangle$\nonumber\\
$D_{s2}^{\ast} \bar D_{s2}^{\ast}$&$\sum_{m,m^{\prime},m_S,m_L}C^{S,m_S}_{2m,2m^{\prime}}C^{J,M}_{Sm_S,Lm_L}
\zeta_{m}^{\mu\nu}\zeta_{m^{\prime}}^{\lambda\rho}|Y_{L\,m_L}\rangle$\\
\bottomrule[1.00pt]
\end{tabular*}.
\end{center}
Here, $C^{e,f}_{ab,cd}$ is the Clebsch-Gordan coefficient, and $|Y_{L\,m_L}\rangle$ stands for the spherical harmonics function. Since the OBE effective potentials contain the tensor forces that couple the $S$ and $D$ partial waves, we need include both $S$-wave and $D$-wave components. For brevity, we list the relevant partial-wave channels $|^{2S+1}L_J\rangle$ of the $D_{s1} \bar D_{s1}$, $D_{s1} \bar D_{s2}^{\ast}$, and $D_{s2}^{\ast} \bar D_{s2}^{\ast}$ systems in the following:
\begin{eqnarray*}
J=0&:&D_{s1} \bar D_{s1}|{}^1{S}_0\rangle/|{}^5{D}_0\rangle,\nonumber\\
&&D_{s2}^{\ast} \bar D_{s2}^{\ast}|{}^1{S}_0\rangle/|{}^5{D}_0\rangle,\nonumber\\
J=1&:&D_{s1} \bar D_{s1}|{}^3{S}_1\rangle/|{}^3{D}_1\rangle,\nonumber\\
&&D_{s1} \bar D_{s2}^{\ast}|{}^3{S}_1\rangle/|{}^{3}{D}_1\rangle/|{}^{5}{D}_1\rangle/|{}^{7}{D}_1\rangle,\nonumber\\
&&D_{s2}^{\ast} \bar D_{s2}^{\ast}|{}^3{S}_1\rangle/|{}^{3}{D}_1\rangle/|{}^{7}{D}_1\rangle,\nonumber\\
J=2&:&D_{s1} \bar D_{s1}|{}^5{S}_2\rangle/|{}^{1}{D}_2\rangle/|{}^{5}{D}_2\rangle,\nonumber\\
&&D_{s1} \bar D_{s2}^{\ast}|^5{S}_2\rangle/|{}^{3}{D}_2\rangle/|{}^{5}{D}_2\rangle/|{}^{7}{D}_2\rangle,\nonumber\\
&&D_{s2}^{\ast} \bar D_{s2}^{\ast}|{}^5{S}_2\rangle/|^{1}{D}_2\rangle/|^{5}{D}_2\rangle/|^{9}{D}_2\rangle,\nonumber\\
J=3&:&D_{s1} \bar D_{s2}^{\ast}|{}^7{S}_3\rangle/|{}^{3}{D}_{3}\rangle/|{}^{5}{D}_{3}\rangle/|{}^{7}{D}_{3}\rangle,\nonumber\\
&&D_{s2}^{\ast} \bar D_{s2}^{\ast}|{}^7{S}_3\rangle/|{}^{3}{D}_3\rangle/|{}^{7}{D}_3\rangle,\nonumber\\
J=4&:&D_{s2}^{\ast} \bar D_{s2}^{\ast}|{}^9{S}_4\rangle/|{}^{5}{D}_4\rangle/|{}^{9}{D}_4\rangle.
\end{eqnarray*}
For each set of quantum numbers $J^{PC}$, the coupled-channel Schr\"odinger equation will be solved in the basis spanned by these partial waves.

\subsubsection{OBE effective potentials}

In the following, we derive the OBE effective potentials of the $T_s \bar T_s$ systems. The derivation consists of three main steps \cite{Chen:2016qju}:
\begin{enumerate}
\item[(i)] For a given transition $AB \to CD$, the scattering amplitude $\mathcal{M}_{E}^{AB\to CD}(\bm{q})$ is obtained from the interaction vertices and the propagator of the exchanged meson $E$:
\begin{eqnarray}
i\mathcal{M}_{E}^{AB \to CD}(\bm{q})=i{\Gamma}^{ACE}\,{P}(q,m_{E})\,i{\Gamma}^{BDE},
\end{eqnarray}
where $\Gamma^{ACE}$ and $\Gamma^{BDE}$ are the interaction vertices derived from the above effective Lagrangians, while ${P}(q,m_E)$ is the propagator of the exchanged meson.
\item[(ii)] Using the Breit approximation \cite{Berestetskii:1982qgu}, the effective potential in the momentum space of the $AB \to CD$ process, represented as $\mathcal{V}_{E}^{AB \to CD}(\bm{q})$, is given by:
\begin{eqnarray}
\mathcal{V}_{E}^{AB \to CD}(\bm{q})=-\frac{\mathcal{M}_{E}^{AB \to CD}(\bm{q})} {4\sqrt{m_{A}m_{B}m_{C}m_{D}}}.
\end{eqnarray}
\item[(iii)] Fourier transform to deduce the effective potential in the coordinate space of the $AB \to CD$ process, denoted as $\mathcal{V}_{E}^{AB \to CD}(\bm{r})$, which is expressed as follows:
\begin{eqnarray}
~~~~~\mathcal{V}_{E}^{AB \to CD}(\bm{r})=\int{\frac{d^3\bm{q}e^{i\bm{q}\cdot\bm{r}}}{(2\pi)^3}}{V}_{E}^{AB \to CD}(\bm{q}){F}_{\rm M}^2(q,\,m_{E}).
\end{eqnarray}
Since the discussed mesons are extended objects, we introduce the monopole form factor ${F}_{\rm M}(q,\,m_{E}) = (\Lambda^2-m_{E}^2)/(\Lambda^2-q^2)$ \cite{Machleidt:1987hj, Epelbaum:2008ga, Esposito:2014rxa, Tornqvist:1993ng, Tornqvist:1993vu} at each interaction vertex to compensate the structure effect of the interaction vertex. Here, $q$ and $m_{{E}}$ are the four momentum and the mass of the exchanged meson, respectively. In the OBE model, $\Lambda$ is the cutoff parameter. The cutoff parameter in the form factor is a free parameter that cannot be precisely determined due to the lack of relevant experimental constraints. Consequently, the sensitivity of the binding energy to $\Lambda$ inevitably limits the absolute predictive power of the model. Nevertheless, drawing on the experience from studies of the deuteron, the hidden-charm pentaquarks $P_c(4312)$, $P_c(4440)$, $P_c(4457)$, and the doubly charmed tetraquark $T_{cc}(3875)$ within the OBE model, a loosely bound state obtained with a cutoff near 1 GeV is generally considered a promising hadronic molecular candidate \cite{Chen:2016qju, Liu:2019zoy, Chen:2022asf}. In light of this, we adopt a cutoff range around 1 GeV in the present work to discuss the bound state properties of the $T_s \bar T_s$ and $T_s T_s$ systems. In this regard, we emphasize that the primary virtue of the OBE model lies in its ability to reveal the existence of hadronic molecular states, whereas its predictive power for the exact numerical value of the binding energy is relatively limited.
\end{enumerate}

By utilising the established principal steps of the OBE model, we can derive the effective potentials in the coordinate space of the $T_s \bar T_s$ systems. To express the OBE effective potentials in a compact form, we introduce a set of operators that act on the spin-orbital wave functions $|^{2S+1}L_J\rangle$ of the $T_s \bar T_s$ systems. These operators encode the polarization contractions of the initial and final mesons, and they are classified according to the discussed systems. For the $D_{s1}\bar D_{s1}$ system, the relevant operators are:
\begin{eqnarray*}
\mathcal{E}_{1}&=&\left({\bm\epsilon^{\dagger}_3}\cdot{\bm\epsilon_1}\right)\left({\bm\epsilon^{\dagger}_4}\cdot{\bm\epsilon_2}\right),\nonumber\\
\mathcal{E}_{2}&=&\left({\bm\epsilon^{\dagger}_3}\times{\bm\epsilon_1}\right)\cdot\left({\bm\epsilon^{\dagger}_4}\times{\bm\epsilon_2}\right),\nonumber\\
\mathcal{F}_{1}&=&S({\bm\epsilon^{\dagger}_3}\times{\bm\epsilon_1},{\bm\epsilon^{\dagger}_4}\times{\bm\epsilon_2},\hat{\bm {r}}),
\end{eqnarray*}
where $S(\bm a,\bm b,\hat{\bm r}) = 3(\hat{\bm r}\cdot\bm a)(\hat{\bm r}\cdot\bm b) - \bm a\cdot\bm b$ is the tensor operator. $\bm{\epsilon}_i$ is the polarization vector of the $D_{s1}$, with the indices $1,2$ ($3,4$) referring to the initial (final) states. Because the $D_{s2}^*$ carries spin 2 and is described by the polarization tensor $\zeta^{\mu\nu}$, the operators involve sums over polarization indices with the Clebsch–Gordan coefficients, and the operator structures become more involved. For the $D_{s1}\bar D_{s2}^*$ system, we define the following operators:
\begin{eqnarray*}
\mathcal{E}_{3}&=&\mathcal{W}_1\left({\bm\epsilon^{\dagger}_3}\cdot{\bm\epsilon_1}\right)\left({\bm\epsilon^{\dagger}_{4a}}\cdot{\bm\epsilon_{2c}}\right)
\left({\bm\epsilon^{\dagger}_{4b}}\cdot {\bm\epsilon_{2d}}\right),\nonumber\\
\mathcal{E}_{3}^{\prime}&=&\mathcal{W}_1\left({\bm\epsilon^{\dagger}_4}\cdot{\bm\epsilon_2}\right)\left({\bm\epsilon^{\dagger}_{3a}}\cdot{\bm\epsilon_{1c}}\right)
\left({\bm\epsilon^{\dagger}_{3b}}\cdot {\bm\epsilon_{1d}}\right),\nonumber\\
\mathcal{E}_{4}&=&\mathcal{W}_1\left({\bm\epsilon^{\dagger}_{4a}}\cdot{\bm\epsilon_{2c}}\right)\left[\left({\bm\epsilon^{\dagger}_{3}}\times{\bm\epsilon_{1}}\right)
\cdot\left({\bm\epsilon^{\dagger}_{4b}}\times
{\bm\epsilon_{2d}}\right)\right],\nonumber\\
\mathcal{E}_{4}^{\prime}&=&\mathcal{W}_1\left({\bm\epsilon^{\dagger}_{3a}}\cdot{\bm\epsilon_{1c}}\right)\left[\left({\bm\epsilon^{\dagger}_{4}}\times{\bm\epsilon_{2}}\right)
\cdot\left({\bm\epsilon^{\dagger}_{3b}}\times {\bm\epsilon_{1d}}\right)\right],\nonumber\\
\mathcal{F}_{2}&=&\mathcal{W}_1\left({\bm\epsilon^{\dagger}_{4a}}\cdot{\bm\epsilon_{2c}}\right)S({\bm\epsilon^{\dagger}_{3}}\times{\bm\epsilon_{1}},{\bm\epsilon^{\dagger}_{4b}}\times {\bm\epsilon_{2d}},\hat{\bm {r}}),\nonumber\\
\mathcal{F}_{2}^{\prime}&=&\mathcal{W}_1\left({\bm\epsilon^{\dagger}_{3a}}\cdot{\bm\epsilon_{1c}}\right)
S({\bm\epsilon^{\dagger}_{4}}\times{\bm\epsilon_{2}},{\bm\epsilon^{\dagger}_{3b}}\times {\bm\epsilon_{1d}},\hat{\bm {r}}),\nonumber\\
\mathcal{E}_{5}&=&\mathcal{W}_1\left({\bm\epsilon^{\dagger}_{3a}}\cdot{\bm\epsilon_1}\right)\left({\bm\epsilon^{\dagger}_{4}}\cdot{\bm\epsilon_{2c}}\right)
\left({\bm\epsilon^{\dagger}_{3b}}\cdot {\bm\epsilon_{2d}}\right),\nonumber\\
\mathcal{E}_{5}^{\prime}&=&\mathcal{W}_1\left({\bm\epsilon^{\dagger}_{4a}}\cdot{\bm\epsilon_2}\right)\left({\bm\epsilon^{\dagger}_{3}}\cdot{\bm\epsilon_{1c}}\right)
\left({\bm\epsilon^{\dagger}_{4b}}\cdot {\bm\epsilon_{1d}}\right),\nonumber\\
\mathcal{F}_{3}&=&\mathcal{W}_1\left({\bm\epsilon^{\dagger}_{3a}}\cdot{\bm\epsilon_1}\right)\left({\bm\epsilon^{\dagger}_{4}}\cdot{\bm\epsilon_{2c}}\right)
S({\bm\epsilon^{\dagger}_{3b}},{\bm\epsilon_{2d}},\hat{\bm {r}}),\nonumber\\
\mathcal{F}_{3}^{\prime}&=&\mathcal{W}_1\left({\bm\epsilon^{\dagger}_{4a}}\cdot{\bm\epsilon_2}\right)\left({\bm\epsilon^{\dagger}_{3}}\cdot{\bm\epsilon_{1c}}\right)
S({\bm\epsilon^{\dagger}_{4b}},{\bm\epsilon_{1d}},\hat{\bm {r}}).
\end{eqnarray*}
Here, the summation over polarization indices is carried out with the weight $\mathcal{W}_1=\sum_{a,b}^{c,d}C^{2,a+b}_{1a,1b}C^{2,c+d}_{1c,1d}$.
For the $D_{s2}^*\bar D_{s2}^*$ system, the operators involve four polarization tensors, which are defined as
\begin{eqnarray*}
\mathcal{E}_{6}&=&\mathcal{W}_2\left({\bm\epsilon^{\dagger}_{3a}}\cdot{\bm\epsilon_{1c}}\right)\left({\bm\epsilon^{\dagger}_{3b}}\cdot{\bm\epsilon_{1d}}\right)
\left({\bm\epsilon^{\dagger}_{4e}}\cdot{\bm\epsilon_{2g}}\right)\left({\bm\epsilon^{\dagger}_{4f}}\cdot{\bm\epsilon_{2h}}\right),\nonumber\\
\mathcal{E}_{7}&=&\mathcal{W}_2\left({\bm\epsilon^{\dagger}_{3a}}\cdot{\bm\epsilon_{1c}}\right)\left({\bm\epsilon^{\dagger}_{4e}}\cdot{\bm\epsilon_{2g}}\right)
\left[\left({\bm\epsilon^{\dagger}_{3b}}\times{\bm\epsilon_{1d}}\right)\cdot\left({\bm\epsilon^{\dagger}_{4f}}\times{\bm\epsilon_{2h}}\right)\right],\nonumber\\
\mathcal{F}_{4}&=&\mathcal{W}_2\left({\bm\epsilon^{\dagger}_{3a}}\cdot{\bm\epsilon_{1c}}\right)\left({\bm\epsilon^{\dagger}_{4e}}\cdot{\bm\epsilon_{2g}}\right)
S({\bm\epsilon^{\dagger}_{3b}}\times{\bm\epsilon_{1d}},{\bm\epsilon^{\dagger}_{4f}}\times{\bm\epsilon_{2h}},\hat{\bm {r}})
\end{eqnarray*}
with $\mathcal{W}_2=\sum_{a,b,c,d}^{e,f,g,h}C^{2,a+b}_{1a,1b}C^{2,c+d}_{1c,1d}C^{2,e+f}_{1e,1f}C^{2,g+h}_{1g,1h}$. To quantitatively evaluate the OBE effective potentials for a given set of quantum numbers $J^{PC}$, the above operators must be evaluated between the spin-orbital wave functions $|^{2S+1}L_J\rangle$ of the $T_s \bar T_s$ systems, and the resulting matrix elements for each operator and each $J$ are collected in Table~\ref{matrix}.

\renewcommand\tabcolsep{0.25cm}
\renewcommand{\arraystretch}{1.50}
\begin{table*}[htbp]
  \caption{The matrix elements for the operators appearing in the OBE effective potentials of the $T_s \bar T_s$ systems. The notation $\operatorname{Diag}(a_1,\dots,a_n)_{[J]}$ indicates a diagonal matrix, while full matrices are shown where necessary.}\label{matrix}
\begin{tabular}{c|c|c}\toprule[1pt]\toprule[1pt]
 Systems & Operators & Matrix elements\\\midrule[1.0pt]
\multirow{5}{*}{$D_{s1}\bar D_{s1}$}&$\mathcal{E}_{1}$
&${\rm{Diag}(1,1)}_{[0]}$,~~~${\rm{Diag}(1,1)}_{[1]}$ ,~~~${\rm{Diag}(1,1,1)}_{[2]}$\\
&$\mathcal{E}_{2}$
&${\rm{Diag}(2,-1)}_{[0]}$,~~~${\rm{Diag}(1,1)}_{[1]}$,~~~${\rm{Diag}(-1,2,-1)}_{[2]}$\\
&$\mathcal{F}_{1}$
&{$\left(\begin{array}{cc}0  &\sqrt{2}\\
                             \sqrt{2}  &2\end{array}\right)_{[0]}$}
                            ,~~~{$\left(\begin{array}{cc}0           &-\sqrt{2}\\
                                     -\sqrt{2}   &1 \end{array}\right)_{[1]}$}
                            ,~~~{$\left(\begin{array}{ccc}
                                0    &\sqrt{\frac{2}{5}}    &-\sqrt{\frac{14}{5}}\\
                                \sqrt{\frac{2}{5}}    &0     &-\frac{2}{\sqrt{7}}\\
                                -\sqrt{\frac{14}{5}}    &-\frac{2}{\sqrt{7}}    &-\frac{3}{7}
                                \end{array}\right)_{[2]}$}\\\hline
\multirow{22}{*}{$D_{s1} \bar D_{s2}^*$}&$\mathcal{E}_{3}$
&${\rm{Diag}(1,1,1,1)}_{[1]}$,~~~${\rm{Diag}(1,1,1,1)}_{[2]}$,~~~${\rm{Diag}(1,1,1,1)}_{[3]}$\\
&$\mathcal{E}_{3}^{\prime}$
&${\rm{Diag}(1,1,1,1)}_{[1]}$,~~~${\rm{Diag}(1,1,1,1)}_{[2]}$,~~~${\rm{Diag}(1,1,1,1)}_{[3]}$\\
&$\mathcal{E}_{4}$
&${\rm{Diag}(\frac{3}{2},\frac{3}{2},\frac{1}{2},-1)}_{[1]}$,~~~${\rm{Diag}(\frac{1}{2},\frac{3}{2},\frac{1}{2},-1)}_{[2]}$,~~~${\rm{Diag}(-1,\frac{3}{2},\frac{1}{2},-1)}_{[3]}$\\
&$\mathcal{E}_{4}^{\prime}$
&${\rm{Diag}(\frac{3}{2},\frac{3}{2},\frac{1}{2},-1)}_{[1]}$,~~~${\rm{Diag}(\frac{1}{2},\frac{3}{2},\frac{1}{2},-1)}_{[2]}$,~~~${\rm{Diag}(-1,\frac{3}{2},\frac{1}{2},-1)}_{[3]}$\\
&$\mathcal{F}_{2}$
&{$\left(\begin{array}{cccc}
                    0 & \frac{3}{5 \sqrt{2}} & \sqrt{\frac{6}{5}} & \frac{\sqrt{\frac{21}{2}}}{5} \\
                    \frac{3}{5 \sqrt{2}} & -\frac{3}{10} & \sqrt{\frac{3}{5}} & -\frac{\sqrt{\frac{3}{7}}}{5} \\
                    \sqrt{\frac{6}{5}} & \sqrt{\frac{3}{5}} & \frac{1}{2} & \frac{2}{\sqrt{35}} \\
                    \frac{\sqrt{\frac{21}{2}}}{5} & -\frac{\sqrt{\frac{3}{7}}}{5} & \frac{2}{\sqrt{35}} & \frac{48}{35}
                   \end{array}\right)_{[1]}$}
                            ,~~~{$\left(\begin{array}{cccc}
                    0  & -\frac{3 \sqrt{2}}{5} & -\sqrt{\frac{7}{10}} & \frac{\sqrt{7}}{5} \\
                    -\frac{3 \sqrt{2}}{5}  & \frac{3}{10} & \frac{3}{\sqrt{35}} & -\frac{3 \sqrt{\frac{2}{7}}}{5} \\
                    -\sqrt{\frac{7}{10}}  & \frac{3}{\sqrt{35}} & -\frac{3}{14} & \frac{4 \sqrt{\frac{2}{5}}}{7} \\
                    \frac{\sqrt{7}}{5}  & -\frac{3 \sqrt{\frac{2}{7}}}{5} & \frac{4 \sqrt{\frac{2}{5}}}{7} & \frac{12}{35}
                   \end{array}\right)_{[2]}$}
                            ,~~~{$\left(\begin{array}{cccc}
                    0 & \frac{3}{5 \sqrt{2}} & -\frac{1}{\sqrt{5}} & -\frac{4 \sqrt{3}}{5} \\
                    \frac{3}{5 \sqrt{2}} & -\frac{3}{35} & -\frac{6 \sqrt{\frac{2}{5}}}{7} & -\frac{6 \sqrt{6}}{35} \\
                    -\frac{1}{\sqrt{5}} & -\frac{6 \sqrt{\frac{2}{5}}}{7} & -\frac{4}{7} & \frac{\sqrt{\frac{3}{5}}}{7} \\
                    -\frac{4 \sqrt{3}}{5} & -\frac{6 \sqrt{6}}{35} & \frac{\sqrt{\frac{3}{5}}}{7} & -\frac{22}{35}
                   \end{array}\right)_{[3]}$}\\
&$\mathcal{F}_{2}^{\prime}$
&{$\left(\begin{array}{cccc}
                    0 & \frac{3}{5 \sqrt{2}} & -\sqrt{\frac{6}{5}} & \frac{\sqrt{\frac{21}{2}}}{5} \\
                    \frac{3}{5 \sqrt{2}} & -\frac{3}{10} & -\sqrt{\frac{3}{5}} & -\frac{\sqrt{\frac{3}{7}}}{5} \\
                    -\sqrt{\frac{6}{5}} & -\sqrt{\frac{3}{5}} & \frac{1}{2} & -\frac{2}{\sqrt{35}} \\
                    \frac{\sqrt{\frac{21}{2}}}{5} & -\frac{\sqrt{\frac{3}{7}}}{5} & -\frac{2}{\sqrt{35}} & \frac{48}{35}
                   \end{array}\right)_{[1]}$}
                             ,~~~{$\left(\begin{array}{cccc}
                    0  & \frac{3 \sqrt{2}}{5} & -\sqrt{\frac{7}{10}} & -\frac{\sqrt{7}}{5} \\
                    \frac{3 \sqrt{2}}{5}  & \frac{3}{10} & -\frac{3}{\sqrt{35}} & -\frac{3 \sqrt{\frac{2}{7}}}{5} \\
                    -\sqrt{\frac{7}{10}} & -\frac{3}{\sqrt{35}} & -\frac{3}{14} & -\frac{4 \sqrt{\frac{2}{5}}}{7} \\
                    -\frac{\sqrt{7}}{5} & -\frac{3 \sqrt{\frac{2}{7}}}{5} & -\frac{4 \sqrt{\frac{2}{5}}}{7} & \frac{12}{35}
                   \end{array}\right)_{[2]}$}
                             ,~~~{$\left(\begin{array}{cccc}
                    0 & \frac{3}{5 \sqrt{2}} & \frac{1}{\sqrt{5}} & -\frac{4 \sqrt{3}}{5} \\
                    \frac{3}{5 \sqrt{2}} & -\frac{3}{35} & \frac{6 \sqrt{\frac{2}{5}}}{7} & -\frac{6 \sqrt{6}}{35} \\
                    \frac{1}{\sqrt{5}} & \frac{6 \sqrt{\frac{2}{5}}}{7} & -\frac{4}{7} & -\frac{\sqrt{\frac{3}{5}}}{7} \\
                    -\frac{4 \sqrt{3}}{5} & -\frac{6 \sqrt{6}}{35} & -\frac{\sqrt{\frac{3}{5}}}{7} & -\frac{22}{35}
                   \end{array}\right)_{[3]}$}\\
&$\mathcal{E}_{5}$
&${\rm{Diag}(\frac{1}{6},\frac{1}{6},\frac{1}{2},1)}_{[1]}$,~~~${\rm{Diag}(\frac{1}{2},\frac{1}{6},\frac{1}{2},1)}_{[2]}$,~~~${\rm{Diag}(1,\frac{1}{6},\frac{1}{2},1)}_{[3]}$\\
&$\mathcal{E}_{5}^{\prime}$
&${\rm{Diag}(\frac{1}{6},\frac{1}{6},\frac{1}{2},1)}_{[1]}$,~~~${\rm{Diag}(\frac{1}{2},\frac{1}{6},\frac{1}{2},1)}_{[2]}$,~~~${\rm{Diag}(1,\frac{1}{6},\frac{1}{2},1)}_{[3]}$\\
&$\mathcal{F}_{3}$
&{$\left(\begin{array}{cccc}
                    0 & -\frac{23}{15 \sqrt{2}} & -2 \sqrt{\frac{2}{15}} & -\frac{\sqrt{\frac{7}{6}}}{5} \\
                    -\frac{23}{15 \sqrt{2}} & \frac{23}{30} & -\frac{2}{\sqrt{15}} & \frac{1}{5 \sqrt{21}} \\
                    2 \sqrt{\frac{2}{15}} & \frac{2}{\sqrt{15}} & \frac{1}{2} & -\frac{2}{\sqrt{35}} \\
                    -\frac{\sqrt{\frac{7}{6}}}{5} & \frac{1}{5 \sqrt{21}} & \frac{2}{\sqrt{35}} & \frac{24}{35}
                   \end{array}\right)_{[1]}$}
                             ,~~~{$\left(\begin{array}{cccc}
                    0  & -\frac{2 \sqrt{2}}{5} & -\sqrt{\frac{7}{10}} & -\frac{\sqrt{7}}{5} \\
                    \frac{2 \sqrt{2}}{5}  & -\frac{23}{30} & -\frac{2}{\sqrt{35}} & \frac{\sqrt{\frac{2}{7}}}{5} \\
                    -\sqrt{\frac{7}{10}}  & \frac{2}{\sqrt{35}} & -\frac{3}{14} & -\frac{4 \sqrt{\frac{2}{5}}}{7} \\
                    \frac{\sqrt{7}}{5}  & \frac{\sqrt{\frac{2}{7}}}{5} & \frac{4 \sqrt{\frac{2}{5}}}{7} & \frac{6}{35}
                   \end{array}\right)_{[2]}$}
                             ,~~~{$\left(\begin{array}{cccc}
                    0 & -\frac{1}{5 \sqrt{2}} & -\frac{1}{\sqrt{5}} & -\frac{2 \sqrt{3}}{5} \\
                    -\frac{1}{5 \sqrt{2}} & \frac{23}{105} & \frac{4 \sqrt{\frac{2}{5}}}{7} & \frac{2 \sqrt{6}}{35} \\
                    \frac{1}{\sqrt{5}} & -\frac{4 \sqrt{\frac{2}{5}}}{7} & -\frac{4}{7} & -\frac{\sqrt{\frac{3}{5}}}{7} \\
                    -\frac{2 \sqrt{3}}{5} & \frac{2 \sqrt{6}}{35} & \frac{\sqrt{\frac{3}{5}}}{7} & -\frac{11}{35}
                   \end{array}\right)_{[3]}$}\\
&$\mathcal{F}_{3}^{\prime}$
&{$\left(\begin{array}{cccc}
                    0 & -\frac{23}{15 \sqrt{2}} & 2 \sqrt{\frac{2}{15}} & -\frac{\sqrt{\frac{7}{6}}}{5} \\
                    -\frac{23}{15 \sqrt{2}} & \frac{23}{30} & \frac{2}{\sqrt{15}} & \frac{1}{5 \sqrt{21}} \\
                    -2 \sqrt{\frac{2}{15}} & -\frac{2}{\sqrt{15}} & \frac{1}{2} & \frac{2}{\sqrt{35}} \\
                    -\frac{\sqrt{\frac{7}{6}}}{5} & \frac{1}{5 \sqrt{21}} & -\frac{2}{\sqrt{35}} & \frac{24}{35}
                   \end{array}\right)_{[1]}$}
                            ,~~~{$\left(\begin{array}{cccc}
                    0 & \frac{2 \sqrt{2}}{5} & -\sqrt{\frac{7}{10}} & \frac{\sqrt{7}}{5} \\
                    -\frac{2 \sqrt{2}}{5} & -\frac{23}{30} & \frac{2}{\sqrt{35}} & \frac{\sqrt{\frac{2}{7}}}{5} \\
                    -\sqrt{\frac{7}{10}} & -\frac{2}{\sqrt{35}} & -\frac{3}{14} & \frac{4 \sqrt{\frac{2}{5}}}{7} \\
                    -\frac{\sqrt{7}}{5} & \frac{\sqrt{\frac{2}{7}}}{5} & -\frac{4 \sqrt{\frac{2}{5}}}{7} & \frac{6}{35}
                   \end{array}\right)_{[2]}$}
                            ,~~~{$\left(\begin{array}{cccc}
                    0 & -\frac{1}{5 \sqrt{2}} & \frac{1}{\sqrt{5}} & -\frac{2 \sqrt{3}}{5} \\
                    -\frac{1}{5 \sqrt{2}} & \frac{23}{105} & -\frac{4 \sqrt{\frac{2}{5}}}{7} & \frac{2 \sqrt{6}}{35} \\
                    -\frac{1}{\sqrt{5}} & \frac{4 \sqrt{\frac{2}{5}}}{7} & -\frac{4}{7} & \frac{\sqrt{\frac{3}{5}}}{7} \\
                    -\frac{2 \sqrt{3}}{5} & \frac{2 \sqrt{6}}{35} & -\frac{\sqrt{\frac{3}{5}}}{7} & -\frac{11}{35}
                   \end{array}\right)_{[3]}$}\\\hline
\multirow{9}{*}{$D_{s2}^{\ast} \bar D_{s2}^{\ast}$}&$\mathcal{E}_{6}$
&${\rm{Diag}(1,1)}_{[0]}$,~~~${\rm{Diag}(1,1,1)}_{[1]}$,~~~${\rm{Diag}(1,1,1,1)}_{[2]}$,~~~${\rm{Diag}(1,1,1)}_{[3]}$,~~~ ${\rm{Diag}(1,1,1)}_{[4]}$\\
&$\mathcal{E}_{7}$
&${\rm{Diag}(\frac{3}{2},\frac{3}{4})}_{[0]}$,~~~${\rm{Diag}(\frac{5}{4},\frac{5}{4},0)}_{[1]}$,~~~${\rm{Diag}(\frac{3}{4},\frac{3}{4},\frac{3}{4},-1)}_{[2]}$,~~~${\rm{Diag}(0,\frac{5}{4},0)}_{[3]}$,~~~${\rm{Diag}(-1,\frac{3}{4},-1)}_{[4]}$\\
&\multirow{5}{*}{$\mathcal{F}_{4}$}
&{$\left(\begin{array}{cc}
                                 0    &\frac{3\sqrt{\frac{7}{10}}}{2}    \\
                                 \frac{3\sqrt{\frac{7}{10}}}{2}      &\frac{15}{14}
                                 \end{array}\right)_{[0]}$}
                            ,~~~{$\left(\begin{array}{ccc}
                                  0 & -\frac{13}{10 \sqrt{2}} & \frac{3 \sqrt{7}}{10} \\
                                  -\frac{13}{10 \sqrt{2}} & \frac{13}{20} & -\frac{3}{5 \sqrt{14}} \\
                                  \frac{3 \sqrt{7}}{10} & -\frac{3}{5 \sqrt{14}} & \frac{36}{35}
                                 \end{array}\right)_{[1]}$}
                             ,~~~{$\left(\begin{array}{cccc}
                                  0 & \frac{3 \sqrt{\frac{7}{2}}}{10} & -\frac{3 \sqrt{\frac{5}{14}}}{2} & \frac{9}{5 \sqrt{14}} \\
                                  \frac{3 \sqrt{\frac{7}{2}}}{10} & 0 & -\frac{3}{2 \sqrt{5}} & 0 \\
                                  -\frac{3 \sqrt{\frac{5}{14}}}{2} & -\frac{3}{2 \sqrt{5}} & -\frac{45}{196} & -\frac{18}{49 \sqrt{5}} \\
                                  \frac{9}{5 \sqrt{14}} & 0 & -\frac{18}{49 \sqrt{5}} & 0
                                 \end{array}\right)_{[2]}$}\\
&&{$\left(\begin{array}{ccc}
                                0    &\frac{3\sqrt{3}}{10}     &-\frac{3\sqrt{3}}{5}\\
                                \frac{3\sqrt{3}}{10}    &\frac{13}{70}      &-\frac{18}{35}\\
                                -\frac{3\sqrt{3}}{5}    &-\frac{18}{35}      &-\frac{33}{70}
                               \end{array}\right)_{[3]}$}
                         ,~~~{$\left(\begin{array}{ccc}
                                0    &\frac{3}{\sqrt{70}}       &-\sqrt{\frac{11}{7}}\\
                               \frac{3}{\sqrt{70}}  &\frac{15}{49}  &-\frac{3\sqrt{\frac{55}{2}}}{49}\\
                                -\sqrt{\frac{11}{7}} &-\frac{3\sqrt{\frac{55}{2}}}{49}  &\frac{65}{98}
                             \end{array}\right)_{[4]}$}\\
\bottomrule[1pt]\bottomrule[1pt]
\end{tabular}
\end{table*}

Using the above operators, the OBE effective potentials of the $T_s \bar T_s$ systems can be obtained. They are expressed in terms of the Yukawa function $Y(m,\Lambda,r)$ and the derivative operators $\mathcal{Z}_r$ and $\mathcal{T}_r$, defined as
\begin{eqnarray*}
&&Y(m,\Lambda,r) = \frac{e^{-mr} - e^{-\Lambda r}}{4\pi r} - \frac{\Lambda^2 - m^2}{8\pi\Lambda} e^{-\Lambda r},\\
&&\mathcal{Z}_r = \frac{1}{r^2}\frac{d}{dr}r^2\frac{d}{dr},~~~~~~~~~~~~~~~
\mathcal{T}_r = r\frac{d}{dr}\frac{1}{r}\frac{d}{dr}.
\end{eqnarray*}
For the cross transition $D_{s1}\bar D_{s2}^* \to D_{s2}^*\bar D_{s1}$, the exchanged momentum carries the energy shift $q_0 = m_{D_{s2}^*}-m_{D_{s1}}$ \cite{Wang:2023ael}, leading to modified arguments $m_0 = \sqrt{m^2-q_0^2}$ and $\Lambda_0 = \sqrt{\Lambda^2-q_0^2}$ in the Yukawa function $Y(m_0,\Lambda_0,r)$. The resulting OBE effective potentials of the $T_s \bar T_s$ systems including the $S$-$D$ wave mixing effect are collected in the following:
\begin{center}
\renewcommand{\arraystretch}{1.50}
\begin{tabular*}{80mm}{@{\extracolsep{\fill}}cc}
\toprule[1.00pt]
\multicolumn{2}{c}{${D}_{s1} \bar {D}_{s1} \to {D}_{s1} \bar {D}_{s1}$ process}\\\hline
~~$f_0$~~ & $-g^{\prime\prime2}_{f_0}\mathcal{E}_{1}Y(m_{f_0},\,\Lambda,\,r)$\\
$\eta$ & $-\frac{25k^2}{162f^2_{\pi}}\left(\mathcal{E}_{2}\mathcal{Z}_r+\mathcal{F}_{1}\mathcal{T}_r\right)Y(m_{\eta},\,\Lambda,\,r)$\\
\multirow{2}{*}{$\phi$}& $-\frac{1}{2}\beta^{\prime\prime2}g^2_V\mathcal{E}_{1}Y(m_{\phi},\,\Lambda,\,r)$\\
&$+\frac{25}{54}\lambda^{\prime\prime2}g^2_V\left(2\mathcal{E}_{2}\mathcal{Z}_r-\mathcal{F}_{1}\mathcal{T}_r\right)Y(m_{\phi},\,\Lambda,\,r)$\\\hline
\multicolumn{2}{c}{${D}_{s1}\bar {D}_{s2}^{\ast}\to {D}_{s1} \bar {D}_{s2}^{\ast}$ process}\\\hline
$f_0$ & $-g^{\prime\prime2}_{f_0}\frac{\mathcal{E}_{3}+\mathcal{E}_{3}^{\prime}}{2}Y(m_{f_0},\,\Lambda,\,r)$\\
$\eta$ & $-\frac{5k^2}{27f^2_{\pi}}\left(\frac{\mathcal{E}_{4}+\mathcal{E}_{4}^{\prime}}{2}\mathcal{Z}_r+\frac{\mathcal{F}_{2}+\mathcal{F}_{2}^{\prime}}{2}\mathcal{T}_r\right)Y(m_{\eta},\,\Lambda,\,r)$\\
\multirow{2}{*}{$\phi$} & $-\frac{1}{2}\beta^{\prime\prime2}g^2_V\frac{\mathcal{E}_{3}+\mathcal{E}_{3}^{\prime}}{2}Y(m_{\phi},\,\Lambda,\,r)$\\
&$+\frac{2}{3}\lambda^{\prime\prime2}g^2_V\left(2\frac{\mathcal{E}_{4}+\mathcal{E}_{4}^{\prime}}{2}\mathcal{Z}_r-\frac{\mathcal{F}_{2}
+\mathcal{F}_{2}^{\prime}}{2}\mathcal{T}_r\right)Y(m_{\phi},\,\Lambda,\,r)$\\\hline
\multicolumn{2}{c}{${D}_{s1} \bar {D}_{s2}^{\ast}\to {D}_{s2}^{\ast} \bar {D}_{s1}$ process}\\\hline
$\eta$ & $-\frac{k^2}{27f^2_{\pi}}\left(\frac{\mathcal{E}_{5}+\mathcal{E}_{5}^{\prime}}{2}\mathcal{Z}_r+\frac{\mathcal{F}_{3}
+\mathcal{F}_{3}^{\prime}}{2}\mathcal{T}_r\right)Y(m_{\eta0},\,\Lambda_0,\,r)$\\
$\phi$ & $\frac{\lambda^{\prime\prime2}g^2_V}{9}\left(2\frac{\mathcal{E}_{5}+\mathcal{E}_{5}^{\prime}}{2}\mathcal{Z}_r-\frac{\mathcal{F}_{3}
+\mathcal{F}_{3}^{\prime}}{2}\mathcal{T}_r\right)Y(m_{\phi0},\,\Lambda_0,\,r)$\\\hline
\multicolumn{2}{c}{${D}_{s2}^{\ast} \bar {D}_{s2}^{\ast} \to {D}_{s2}^{\ast} \bar {D}_{s2}^{\ast}$ process}\\\hline
$f_0$ & $-g^{\prime\prime2}_{f_0}\mathcal{E}_{6}Y(m_{f_0},\,\Lambda,\,r)$\\
$\eta$ & $-\frac{2k^2}{9f^2_{\pi}}\left(\mathcal{E}_{7}\mathcal{Z}_r
+\mathcal{F}_{4}\mathcal{T}_r\right)Y(m_{\eta},\,\Lambda,\,r)$\\
\multirow{2}{*}{$\phi$} & $-\frac{1}{2}\beta^{\prime\prime2}g^2_V\mathcal{E}_{6}Y(m_{\phi},\,\Lambda,\,r)$\\
&$+\frac{2}{3}\lambda^{\prime\prime2}g^2_V\left(2\mathcal{E}_{7}\mathcal{Z}_r-\mathcal{F}_{4}\mathcal{T}_r\right)Y(m_{\phi},\,\Lambda,\,r)$\\
\bottomrule[1.00pt]
\end{tabular*}
\end{center}
After summing over all exchanged mesons, the total effective potentials for each system in the coordinate space can be expressed as:
\begin{equation}
\mathcal{V}_{{\rm Total}}(r) = \mathcal{V}_{f_0}(r)+\mathcal{V}_{\eta}(r)+\mathcal{V}_{\phi}(r).
\end{equation}

\begin{figure}[htbp]
  \centering
  \includegraphics[scale=0.39]{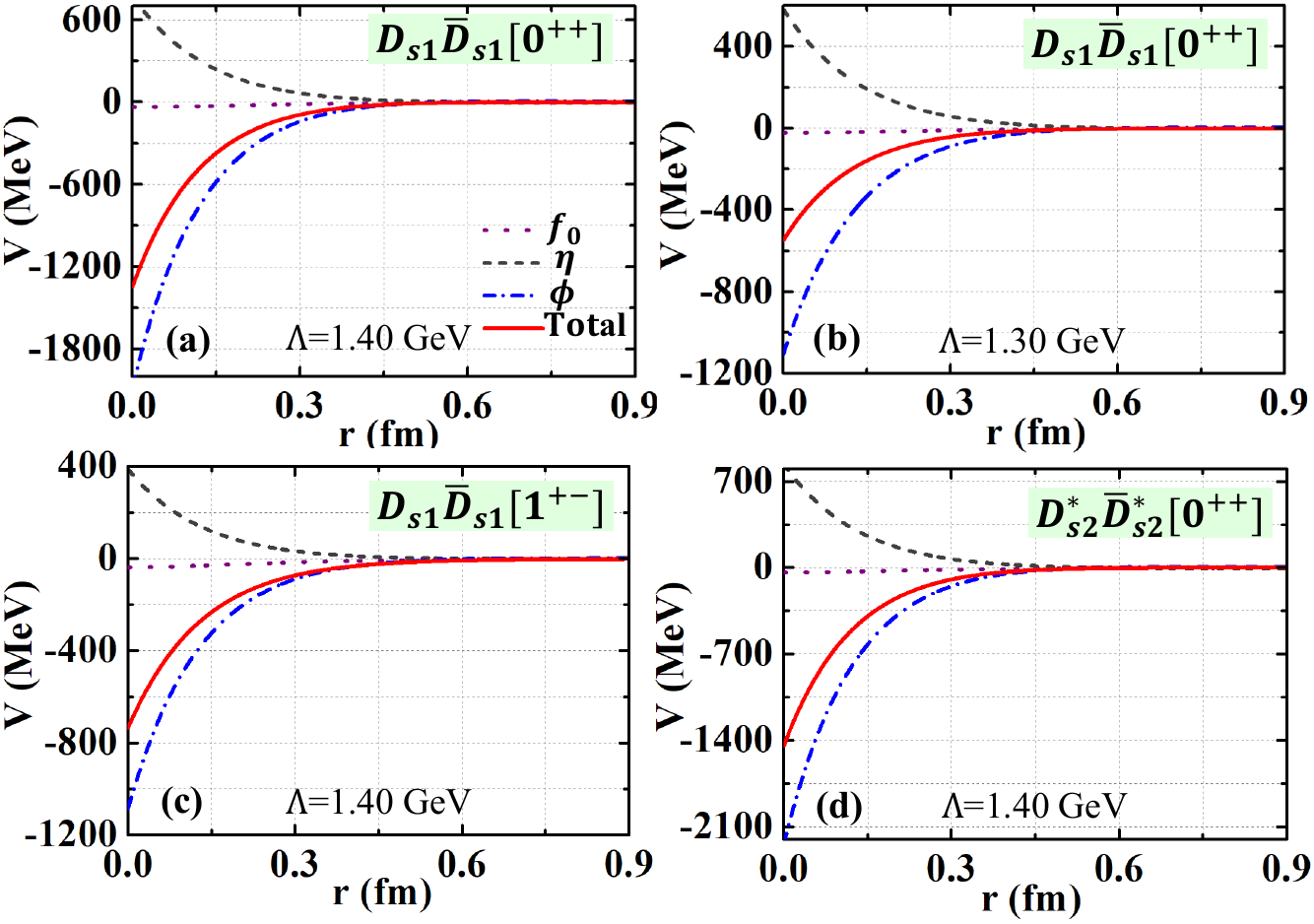}
  \caption{The OBE effective interactions for several typical systems.}\label{Potential}
\end{figure}

In Fig. \ref{Potential}, we display the individual contributions from the $f_0$, $\eta$, and $\phi$ exchanges to the total effective potential for several typical systems. It is clear that the $\phi$ exchange provides the dominant attraction, whereas the $\eta$ exchange produces a substantial repulsive component that partially cancels this attraction. The $f_0$ exchange, in contrast, gives only a minor attractive contribution. A comparison between panels (a) and (b) indicates that these features are essentially insensitive to the choice of the cutoff parameter. Similarly, panels (a) and (c) demonstrate that the same pattern persists when the $J^{PC}$ quantum numbers are varied within the same system. Moreover, panels (a) and (d) reveal that this behavior is also observed for different systems sharing the same $J^{PC}$ quantum numbers. Consequently, the above conclusions appear to be general for all systems considered in this work.

\subsection{Bound state solutions of the $T_s \bar T_s$ systems}

In the following, we analyze the binding properties of the $D_{s1}\bar D_{s1}$, $D_{s1}\bar D_{s2}^*$, and $D_{s2}^*\bar D_{s2}^*$ systems. In the numerical analysis, we attempt to find the loosely bound state solutions for the discussed systems by varying the cutoff parameter $\Lambda$ over a discrete mesh with a step size of $0.01\,\text{GeV}$. For each loosely bound state, we present typical results at three representative values of $\Lambda$ to illustrate the dependence. These values are not arbitrary, and they are chosen to bracket the region where the bound state is shallow and to demonstrate how the properties evolve with the cutoff. Here, we emphasize that once the cutoff parameter reaches a certain value, the binding energy of the discussed system starts to increase from zero, and as the cutoff parameter is continuously increased, the binding energy also increases continuously. This continuous behavior is physically expected. The binding energy $E$ is taken directly as the eigenvalue, with negative values indicating a bound state, while the spatial wave functions $\phi_i(r)$ for each channel $i$ are obtained from the eigenvector components. From the spatial wave functions, we further discuss two physical quantities that characterize the loosely bound state:
\begin{itemize}
\item[(i)] The RMS radius, defined by
\begin{eqnarray*}
r_{\rm RMS} = \sqrt{\sum_i \int_0^\infty r^2 \phi_i(r)^2 r^2 dr},
\end{eqnarray*}
which quantifies the spatial extension of the bound state. A large $r_{\rm RMS}$ signals a loosely bound state, similar to the deuteron \cite{Chen:2016qju}.
\item[(ii)] The probability of each channel $i$, defined by
\begin{eqnarray*}
P(i) = \int_0^\infty \phi_i(r)^2 r^2 dr,
\end{eqnarray*}
which indicates the dominant composition of the loosely bound state.
\end{itemize}

In the following, we present the bound state properties for each $J^{PC}$ quantum-numbers of the $D_{s1}\bar D_{s1}$, $D_{s1}\bar D_{s2}^*$, and $D_{s2}^*\bar D_{s2}^*$ systems. For each loosely bound state, we examine the cutoff $\Lambda$ dependence of the binding energy $E$, mass $M$, RMS radius $r_{\rm RMS}$, and partial-wave probabilities $P(i)$. These physical quantities collectively determine whether the discussed system qualifies as promising hadronic molecular candidate.

\subsubsection{The $D_{s1}\bar D_{s1}$ system}

We now analyze the bound state properties of the $D_{s1}\bar D_{s1}$ system for three $J^{PC}$ quantum numbers: $0^{++}$, $1^{+-}$, and $2^{++}$. The corresponding numerical results, obtained from the coupled-channel Schr\"odinger equation including the $S$-$D$ wave mixing effect, are listed in Table \ref{massspectra1}.

\renewcommand\tabcolsep{0.21cm}
\renewcommand{\arraystretch}{1.50}
\begin{table}[!htbp]
\caption{Bound state properties of the $D_{s1}\bar D_{s1}$ system. The cutoff $\Lambda$, binding energy $E$, mass $M$, RMS radius $r_{\rm RMS}$, and partial-wave probabilities $P(i)$ are given in units of GeV, MeV, MeV, fm, and $\%$, respectively.}\label{massspectra1}
\begin{tabular}{ccccccc}\toprule[1pt]\toprule[1pt]
\multicolumn{7}{c}{$0^{++}$}\\\hline
$\Lambda$ &$E$ &$M$ &$r_{\rm RMS}$ &$P$(${}^1{S}_{0})$&$P$(${}^5{D}_{0})$&\\
1.44&$-1.06$&5061.36&2.81& 99.00&1.00&\\
1.46&$-5.83$&5056.59&1.28&98.06&1.94&\\
1.48&$-14.70$&5047.72&0.85&97.49&2.51&\\\hline
\multicolumn{7}{c}{$1^{+-}$}\\\hline
$\Lambda$ &$E$ &$M$ &$r_{\rm RMS}$ &$P$(${}^3{S}_{1})$&$P$(${}^3{D}_{1})$&\\
1.50&$-0.62$&5061.80&3.59&99.14&0.86&\\
1.54&$-6.33$&5056.09&1.26&97.61&2.39&\\
1.57&$-14.94$&5047.48&0.87&96.74&3.26&\\\hline
\multicolumn{7}{c}{$2^{++}$}\\\hline
$\Lambda$ &$E$ &$M$ &$r_{\rm RMS}$ &$P$(${}^5{S}_{2})$&$P$(${}^1{D}_{2})$&$P$(${}^5{D}_{2})$\\
1.61&$-0.54$&5061.88&3.87&98.59&0.25&1.16\\
1.69&$-5.00$&5057.42&1.51&95.56&0.84&3.61\\
1.77&$-14.41$&5048.01&1.01&92.35&1.54&6.10\\
\bottomrule[1pt]\bottomrule[1pt]
\end{tabular}
\end{table}

Table~\ref{massspectra1} reveals several systematic features. For each $J^{PC}$ channel, a loosely bound state with a binding energy of order $1$~MeV and an RMS radius of several fermis appears at a critical cutoff $\Lambda$. As $\Lambda$ increases, the binding energy grows monotonically, the RMS radius shrinks, and the $S$-wave dominance remains above $92\%$ in all cases, with the $D$-wave admixtures gradually increasing. Among the three channels, the $0^{++}$ state becomes bound at the smallest cutoff and develops the deepest binding for a given $\Lambda$, indicating the strongest attraction, while the $2^{++}$ state requires the largest cutoff, reflecting weaker attraction at higher $J$.

Based on the above analysis, the $D_{s1}\bar D_{s1}$ system yields three promising candidates of the hidden-charm hidden-strangeness molecular tetraquarks with the quantum numbers $J^{PC}=0^{++}$, $1^{+-}$, and $2^{++}$. Their predicted masses lie in the range $5.04$–$5.06$~GeV, just below the $D_{s1}\bar D_{s1}$ threshold.

\renewcommand\tabcolsep{0.60cm}
\renewcommand{\arraystretch}{1.50}
\begin{table}[!htbp]
\centering
\caption{Binding energies (in MeV) of the $D_{s1}\bar D_{s1}$ states with $J^{PC}=0^{++}$ and $1^{+-}$, obtained with the same cutoff parameter $\Lambda$.}\label{samecutoff}
\begin{tabular}{ccc}\toprule[1pt]\toprule[1pt]
$\Lambda\,({\rm MeV})$ &$D_{s1}\bar D_{s1}[0^{++}]$ &$D_{s1}\bar D_{s1}[1^{+-}]$\\
1490&$-20.73$&$-0.12$\\
1500&$-27.87$&$-0.62$\\
1510&$-36.11$&$-1.46$\\
\bottomrule[1pt]\bottomrule[1pt]
\end{tabular}
\end{table}

In the above discussion, we investigated the bound-state properties of the $D_{s1}\bar{D}_{s1}$ system by scanning the cutoff parameter, with the results presented in Table \ref{massspectra1}. In the following, we take the $J^{PC}=0^{++}$ and $1^{+-}$ $D_{s1}\bar{D}_{s1}$ states as examples and discuss their binding energies with the same cutoff parameter, which leads to several interesting conclusions. The corresponding numerical results are listed in Table \ref{samecutoff}. From Table \ref{samecutoff}, one can see that, when the same cutoff parameter is adopted, the $J^{PC}=0^{++}$ $D_{s1}\bar{D}_{s1}$ state exhibits stronger attraction, deeper binding, and a smaller mass compared with the $J^{PC}=1^{+-}$ $D_{s1}\bar{D}_{s1}$ state. This comparison is crucial for discussing the level ordering of these hadronic molecular states and can provide valuable guidance for future experimental searches.

\subsubsection{The $D_{s1}\bar D_{s2}^*$ system}

\renewcommand\tabcolsep{0.47cm}
\renewcommand{\arraystretch}{1.50}
\begin{table}[!htbp]
\caption{Bound state properties of the $D_{s1} \bar D_{s2}^*$ system. Conventions are the same as Table~\ref{massspectra1}.}\label{massspectra2}
\begin{tabular}{ccccc}\toprule[1pt]\toprule[1pt]
\multicolumn{5}{c}{$1^{++}$}\\\hline
$\Lambda$ &$E$ &$M$ &$r_{\rm RMS}$ &$P$(${}^3{S}_{1})$\\
1.46&$-0.69$&5099.62&3.37&99.53\\
1.49&$-7.68$&5092.63&1.12&98.83\\
1.51&$-16.86$&5083.45&0.79&98.58\\\hline
\multicolumn{5}{c}{$1^{+-}$}\\\hline
$\Lambda$ &$E$ &$M$ &$r_{\rm RMS}$ &$P$(${}^3{S}_{1})$\\
1.46&$-0.40$&5099.91&4.13&99.78\\
1.49&$-7.26$&5093.05&1.14&99.31\\
1.51&$-16.85$&5083.46&0.78&99.15\\\hline
\multicolumn{5}{c}{$2^{++}$}\\\hline
$\Lambda$ &$E$ &$M$ &$r_{\rm RMS}$ &$P$(${}^5{S}_{2})$\\
1.54&$-0.33$&5099.98&4.45&99.55\\
1.59&$-5.40$&5094.91&1.35&98.46\\
1.63&$-14.30$&5086.01&0.89&97.78\\\hline
\multicolumn{5}{c}{$2^{+-}$}\\\hline
$\Lambda$ &$E$ &$M$ &$r_{\rm RMS}$ &$P$(${}^5{S}_{2})$\\
1.61&$-0.44$&5099.87&4.07&99.70\\
1.66&$-4.08$&5096.23&1.54&99.14\\
1.72&$-13.82$&5086.49&0.90&98.559\\\hline
\multicolumn{5}{c}{$3^{++}$}\\\hline
$\Lambda$ &$E$ &$M$ &$r_{\rm RMS}$ &$P$(${}^7{S}_{3})$\\
1.67&$-0.55$&5099.76&3.85&98.79\\
1.77&$-5.27$&5095.04&1.50&96.07\\
1.87&$-15.23$&5085.08&1.01&93.16\\\hline
\multicolumn{5}{c}{$3^{+-}$}\\\hline
$\Lambda$ &$E$ &$M$ &$r_{\rm RMS}$ &$P$(${}^7{S}_{3})$\\
1.60&$-0.48$&5099.83&4.04&98.76\\
1.69&$-5.42$&5094.89&1.46&95.78\\
1.77&$-14.55$&5085.76&1.00&93.14\\
\bottomrule[1pt]\bottomrule[1pt]
\end{tabular}
\end{table}

We now proceed to analyze the bound state properties of the $D_{s1}\bar D_{s2}^*$ system. The numerical results for six $J^{PC}$ quantum numbers ($1^{++}$, $1^{+-}$, $2^{++}$, $2^{+-}$, $3^{++}$, and $3^{+-}$) are presented in Table~\ref{massspectra2}. The table shows that for each $J^{PC}$ channel, a shallow bound state appears at a cutoff in the range $\Lambda \simeq 1.46$--$1.67$~GeV. As in the $D_{s1}\bar D_{s1}$ case, increasing $\Lambda$ systematically deepens the binding and reduces the RMS radius. In all cases, the $S$-wave component remains above $93\%$, reflecting the dominant role of the $S$-wave interaction. A notable feature of this system is the near degeneracy of the binding properties for states with opposite $C$-parity at a given $J$, especially for the $J=1$ doublet. This suggests that experimental searches should observe $C$-parity doublets, providing a clean signature for the molecular interpretation.

The $J^{PC}=2^{+-}$ state deserves special attention, as this spin-parity combination is exotic and cannot be realized by a conventional meson. Its observation would provide unambiguous evidence for an exotic hadron. Except for the $J=1$ and $2$ molecules, our analysis reveals that the system also supports the loosely bound states with higher total angular momentum, notably $J=3$. Their discovery would provide the first examples of the hadronic molecules with spin 3.

A clear ordering of the binding strength with total angular momentum $J$ is observed: the $J=1$ channels become bound at the smallest $\Lambda$ and develop the deepest binding, while the $J=2$ and $3$ channels require progressively larger $\Lambda$ to achieve comparable binding. This trend reflects the decreasing attraction with increasing $J$, due to the interplay of the spin-dependent forces.

Based on the above analysis, the $D_{s1}\bar D_{s2}^*$ states with $J^{PC}=1^{++}$, $1^{+-}$, $2^{++}$, $2^{+-}$, $3^{++}$, and $3^{+-}$ are promising candidates of the hidden-charm hidden-strangeness molecular tetraquarks. We strongly recommend experimental searches for these $D_{s1}\bar D_{s2}^*$ molecular tetraquarks near $5.08$–$5.10$~GeV, slightly below the $D_{s1}\bar D_{s2}^*$ threshold.

\subsubsection{The $D_{s2}^*\bar D_{s2}^*$ system}

We now investigate the bound state properties of the $D_{s2}^*\bar D_{s2}^*$ system. The numerical results for five quantum numbers ($0^{++}$, $1^{+-}$, $2^{++}$, $3^{+-}$, and $4^{++}$) are presented in Table \ref{massspectra3}.

\renewcommand\tabcolsep{0.12cm}
\renewcommand{\arraystretch}{1.50}
\begin{table}[!htbp]
\caption{Bound state properties of the $D_{s2}^* \bar D_{s2}^*$ system. Conventions are the same as Table~\ref{massspectra1}.}\label{massspectra3}
\begin{tabular}{cccccccc}\toprule[1pt]\toprule[1pt]
\multicolumn{8}{c}{$0^{++}$}\\\hline
$\Lambda$ &$E$ &$M$ &$r_{\rm RMS}$ &$P$(${}^1{S}_{0})$&$P$(${}^5{D}_{0})$\\
1.42&$-1.28$&5136.92&2.57&98.61&1.39\\
1.44&$-6.54$&5131.66&1.21&97.43&2.57 \\
1.46&$-16.22$&5121.98&0.82&96.72&3.28 \\\hline
\multicolumn{8}{c}{$1^{+-}$}\\\hline
$\Lambda$ &$E$ &$M$ &$r_{\rm RMS}$ &$P$(${}^3{S}_{1})$&$P$(${}^3{D}_{1})$&$P$(${}^7{D}_{1})$\\
1.43&$-0.46$&5137.74&3.95&99.10&0.49&0.41\\
1.46&$-6.50$&5131.7&1.22&97.35&1.43&1.22\\
1.48&$-14.75$&5123.45&0.86&96.62&1.82&1.56\\\hline
\multicolumn{8}{c}{$2^{++}$}\\\hline
$\Lambda$ &$E$ &$M$ &$r_{\rm RMS}$ &$P$(${}^5{S}_{2})$&$P$(${}^1{D}_{2})$&$P$(${}^5{D}_{2})$&$P$(${}^7{D}_{2})$\\
1.47&$-0.36$&5137.84&4.33&99.15&0.21&0.45&0.19\\
1.51&$-5.78$&5132.42&1.31&97.14&0.72&1.49&0.65\\
1.54&$-14.53$&5123.67&0.88&96.02&1.02&2.06&0.90\\\hline
\multicolumn{8}{c}{$3^{+-}$}\\\hline
$\Lambda$ &$E$ &$M$ &$r_{\rm RMS}$ &$P$(${}^7{S}_{3})$&$P$(${}^3{D}_{3})$&$P$(${}^7{D}_{3})$\\
1.56&$-0.51$&5137.69&3.87&98.85&0.31&0.83\\
1.62&$-4.84$&5133.36&1.47&96.57&0.98&2.45\\
1.68&$-13.86$&5124.34&0.96&94.37&1.70&3.93\\\hline
\multicolumn{8}{c}{$4^{++}$}\\\hline
$\Lambda$ &$E$ &$M$ &$r_{\rm RMS}$ &$P$(${}^9{S}_{4})$&$P$(${}^5{D}_{4})$&$P$(${}^9{D}_{4})$\\
1.62&$-0.66$&5137.54&3.58&98.54&0.19&1.27\\
1.71&$-5.53$&5132.67&1.47&95.66&0.63&3.72\\
1.79&$-14.09$&5124.11&1.03&93.03&1.10&5.87\\
\bottomrule[1pt]\bottomrule[1pt]
\end{tabular}
\end{table}

The results for the $D_{s2}^*\bar D_{s2}^*$ system follow the same general trends observed in the previous systems. For each $J^{PC}$ channel, a loosely bound state appears at a cutoff $\Lambda$ ranging from $1.42$ to $1.62$~GeV. The binding energy increases monotonically with $\Lambda$, the RMS radius decreases correspondingly, and the $S$-wave component remains dominant (above $93\%$). The $J^{PC}=4^{++}$ state is of particular interest because it carries the highest total angular momentum among all candidates considered in this work. High-spin molecular states are currently missing in experiments \cite{ParticleDataGroup:2024cfk}. Its discovery would provide the first example of a hadronic molecule with spin 4, and therefore experimental searches should be prioritized.

From the preceding analysis, the $D_{s2}^*\bar D_{s2}^*$ system yields promising candidates of hidden-charm hidden-strangeness molecular tetraquarks in the $J^{PC}=0^{++}$, $1^{+-}$, $2^{++}$, $3^{+-}$, and $4^{++}$ channels. We strongly encourage experimental searches for these $D_{s2}^*\bar D_{s2}^*$ molecular tetraquarks near $5.12$–$5.14$~GeV.

According to the above systematic analysis of the bound state properties of the $D_{s1}\bar D_{s1}$, $D_{s1}\bar D_{s2}^*$, and $D_{s2}^*\bar D_{s2}^*$ systems, we have established the characteristic spectrum of promising hidden-charm hidden-strangeness molecular tetraquark candidates arising from the $T$-doublet charmed-strange mesons and their antiparticles. Specifically, the following states are identified as promising candidates of hidden-charm hidden-strangeness molecular tetraquarks:
\begin{itemize}
\item $D_{s1}\bar D_{s1}$ states with $J^{PC}=0^{++}$, $1^{+-}$, and $2^{++}$,
\item $D_{s1}\bar D_{s2}^*$ states with $J^{PC}=1^{+\pm}$, $2^{+\pm}$, and $3^{+\pm}$,
\item $D_{s2}^*\bar D_{s2}^*$ states with $J^{PC}=0^{++}$, $1^{+-}$, $2^{++}$, $3^{+-}$, and $4^{++}$.
\end{itemize}
Among these, two categories deserve special emphasis:
(i) The $D_{s1}\bar D_{s2}^*$ state with $J^{PC}=2^{+-}$ carries exotic spin-parity quantum numbers that cannot be realized by conventional $c\bar c$ mesons. Its observation would provide unambiguous evidence for an exotic hadron. (ii) Our analysis supports the existence of the loosely bound states with the higher total angular momentum, specifically $J\ge 3$. Among them, the $D_{s2}^*\bar D_{s2}^*$ state with $J^{PC}=4^{++}$ possesses total angular momentum $J=4$, the highest among all candidates. Such a high-spin hadronic molecule is a rare entity, and its discovery would represent a milestone in the spectroscopy of high-spin exotic hadrons.

\section{Doubly-charm doubly-strangeness molecular $T_s T_s$ tetraquarks}\label{sec3}

Having systematically identified promising candidates of hidden-charm hidden-strangeness molecular tetraquarks in the $T_s\bar T_s$ systems, we now extend our analysis to doubly-charm doubly-strangeness $T_s T_s$ systems. These systems consist of two $T$-doublet charmed-strange mesons, carrying open charm $C=+2$ and open strangeness $S=+2$, and are therefore flavor-exotic. For the $S$-wave configurations, the allowed quantum numbers are $J^P = 0^+,\,2^+$ for the $D_{s1}D_{s1}$ system, $J^P = 0^+,\,1^+,\,2^+$ for the $D_{s1}D_{s2}^*$ system, and $J^P = 0^+,\, 2^+,\,4^+$ for the $D_{s2}^*D_{s2}^*$ system. Thus, the $T_s T_s$ systems serve as a natural laboratory for high-spin doubly-charm doubly-strangeness exotic hadrons. Among these, we have identified several loosely bound states, whose properties are summarized in Table~\ref{massspectra4}.

\renewcommand\tabcolsep{0.13cm}
\renewcommand{\arraystretch}{1.50}
\begin{table}[!htbp]
\caption{Bound state properties of the $T_s T_s$ systems. Conventions are the same as Table~\ref{massspectra1}. Here, we note that the channels with probabilities below $5\times 10^{-3}$\% are denoted simply as 0.}\label{massspectra4}
\begin{tabular}{cccccccc}\toprule[1pt]\toprule[1pt]
\multicolumn{8}{c}{$D_{s1} D_{s1}$ state with $J^P=2^+$}\\\hline
$\Lambda$ &$E$ &$M$ &$r_{\rm RMS}$ &$P$(${}^5{S}_{2})$&$P$(${}^1{D}_{2})$&$P$(${}^5{D}_{2})$\\
1.61&$-1.63$&5068.79&2.09&99.88&0.01&0.11\\
1.63&$-8.01$&5062.41&0.95&99.76&0.02&0.22\\
1.64&$-12.57$&5057.85&0.77&99.71&0.02&0.27\\\hline
\multicolumn{8}{c}{$D_{s1} D_{s2}^*$ state with $J^P=3^+$}\\\hline
$\Lambda$ &$E$ &$M$ &$r_{\rm RMS}$ &$P$(${}^7{S}_{3})$&$P$(${}^3{D}_{3})$&$P$(${}^5{D}_{3})$&$P$(${}^7{D}_{3})$\\
1.46&$-1.51$&5102.80&2.18&99.99&0&0&0.01\\
1.48&$-8.38$&5095.93&0.95&99.98&0&0&0.02\\
1.49&$-13.49$&5090.82&0.76&99.97&0&0&0.03\\\hline
\multicolumn{8}{c}{$D_{s2}^* D_{s2}^*$ state with $J^P=4^+$}\\\hline
$\Lambda$ &$E$ &$M$ &$r_{\rm RMS}$ &$P$(${}^9{S}_{4})$&$P$(${}^5{D}_{4})$&$P$(${}^9{D}_{4})$\\
1.45&$-0.24$&5137.96&4.58&99.99&0&0.01\\
1.47&$-5.15$&5133.05&1.19&99.97&0&0.03\\
1.49&$-14.78$&5123.42&0.73&99.95&0&0.05\\
\bottomrule[1pt]\bottomrule[1pt]
\end{tabular}
\end{table}

Table \ref{massspectra4} reveals several common features across the $T_s T_s$ systems. In each case, a loosely bound state appears at a cutoff $\Lambda$ around $1.4$-$1.6$ GeV, with binding energy of order $1$~MeV and an extended RMS radius of several fermis. As $\Lambda$ increases, the binding energy grows monotonically and the RMS radius shrinks, following the same trend observed in the $T_s\bar T_s$ systems. A striking feature of the $T_s T_s$ bound states is their extremely pure $S$-wave nature: the $S$-wave probability exceeds $99.7\%$ in all cases, with the $D$-wave admixtures being negligibly small (typically less than $0.3\%$). This is in contrast to the $T_s\bar T_s$ systems, where the $D$-wave mixing, while still small, is somewhat more pronounced. The purity of the $S$-wave configurations in the $T_s T_s$ systems suggests that the tensor forces play a less significant role in the doubly-charm doubly-strangeness systems compared with their hidden-charm hidden-strangeness counterparts.

Several specific features of the $T_s T_s$ systems are worth highlighting. For the $D_{s1}D_{s2}^*$ system, although the $S$-wave allows total angular momenta $J=1,2,3$, only the highest spin $J=3$ channel is found to be bound. This selectivity reflects the spin-dependent nature of the light-meson exchange interactions. The $D_{s2}^*D_{s2}^*$ system yields a $J^P=4^+$ bound state with the highest total angular momentum among all candidates considered in this work. Such high-spin molecular states are rarely predicted in the hadronic molecule spectroscopy \cite{Chen:2016qju, Liu:2019zoy, Chen:2022asf}, and their discovery would provide unprecedented evidence for the high-spin exotic hadrons.

Based on the preceding analysis of the bound state properties of the $D_{s1}D_{s1}$, $D_{s1}D_{s2}^*$, and $D_{s2}^*D_{s2}^*$ systems, we have established the characteristic spectrum of promising doubly-charm doubly-strangeness molecular tetraquark candidates arising from two $T$-doublet charmed-strange mesons. These states are absolutely flavor-exotic, they contain two charm quarks and two strange anti-quarks ($cc\bar s\bar s$), a combination that cannot be accommodated in conventional hadrons. Their observation would thus provide unambiguous evidence for genuine multiquark exotic states. The identified doubly-charm doubly-strangeness molecular tetraquark candidates are:
\begin{itemize}
\item $D_{s1}D_{s1}$ state with $J^P=2^{+}$,
\item $D_{s1}D_{s2}^*$ state with $J^P=3^{+}$,
\item $D_{s2}^*D_{s2}^*$ state with $J^P=4^{+}$.
\end{itemize}
Among these, the $J=3$ and $4$ states are particularly noteworthy because the high-spin molecules are currently missing in experiments \cite{ParticleDataGroup:2024cfk}, and their discovery would open a new window into the spectroscopy of high-spin exotic hadrons.

Finally, we briefly discuss the influence of the coupled-channel effect on the binding properties of the $T_s\bar T_s$ and $T_s T_s$ systems. In the preceding analysis, we adopted the single-channel approximation including the $S$-$D$ wave mixing effect to investigate the binding properties of molecular tetraquarks formed by the $T$-doublet charmed-strange mesons and their antiparticles, as well as by two $T$-doublet charmed-strange mesons. To assess the robustness of our conclusions, we have also performed the coupled-channel calculations that incorporate all relevant channels with nearby thresholds. Our coupled-channel results show that the primary effect of the channel coupling is a redistribution of probabilities among the constituent channels. However, no additional bound states appear beyond those already identified in the single-channel analysis. Therefore, the single-channel treatment with the $S$-$D$ wave mixing effect captures the essential features of the binding properties of the $T_s\bar T_s$ and $T_s T_s$ tetraquark systems, and the corresponding molecular candidates listed in this work are robust against coupled-channel corrections.

\section{Discussion and Conclusion}\label{sec4}

In this work, we perform a systematic investigation of the mass spectrum of doubly-charm molecular tetraquarks with strangeness $S=0$ or $2$, formed from the $T$-doublet charmed-strange mesons. Two distinct families are considered: hidden-charm hidden-strangeness $T_s\bar T_s$ systems and doubly-charm doubly-strangeness $T_s T_s$ systems. Using the OBE model with the effective Lagrangians that respect the heavy-quark, chiral, and hidden-gauge symmetries, we derive the interaction potentials of the $T_s\bar T_s$ systems mediated by $f_0$, $\eta$, and $\phi$ exchanges. We solve the coupled-channel Schr\"odinger equation and extract the binding energies, masses, RMS radii, and partial-wave compositions for each set of $J^{PC}$ quantum numbers. Our analysis reveals a rich spectrum of doubly-charm molecular tetraquarks with strangeness $S=0$ or $2$ formed from the $T$-doublet charmed-strange mesons.

For the hidden-charm hidden-strangeness $T_s\bar T_s$ tetraquark systems, the promising candidates are the $D_{s1}\bar D_{s1}$ states with $J^{PC}=0^{++},\,1^{+-},\,2^{++}$, the $D_{s1}\bar D_{s2}^*$ states with $J^{PC}=1^{++},\,1^{+-},\,2^{++},\,2^{+-},\,3^{++},\,3^{+-}$, and the $D_{s2}^*\bar D_{s2}^*$ states with $J^{PC}=0^{++},\,1^{+-},\,2^{++},\,3^{+-},\,4^{++}$. For the doubly-charm doubly-strangeness $T_s T_s$ tetraquark systems, which are absolutely flavor-exotic ($cc\bar s\bar s$), we identify the following candidates: the $D_{s1}D_{s1}$ state with $J^P=2^+$, the $D_{s1}D_{s2}^*$ state with $J^P=3^+$, and the $D_{s2}^*D_{s2}^*$ state with $J^P=4^+$. Among these, the $D_{s1}\bar D_{s2}^*$ state with $J^{PC}=2^{+-}$ carries exotic spin-parity quantum numbers strictly forbidden for conventional $c\bar c$ mesons, providing a clean experimental signature that would unambiguously distinguish a genuine exotic hadron. Moreover, the $D_{s2}^*\bar D_{s2}^*$ state with $J^{PC}=4^{++}$ and the $D_{s2}^*D_{s2}^*$ state with $J^P=4^+$ are extremely rare in hadronic molecule spectroscopy, and their discovery would open a new window into the study of high-spin exotic hadrons.

\begin{figure}[htbp]
\centering
\includegraphics[scale=0.43]{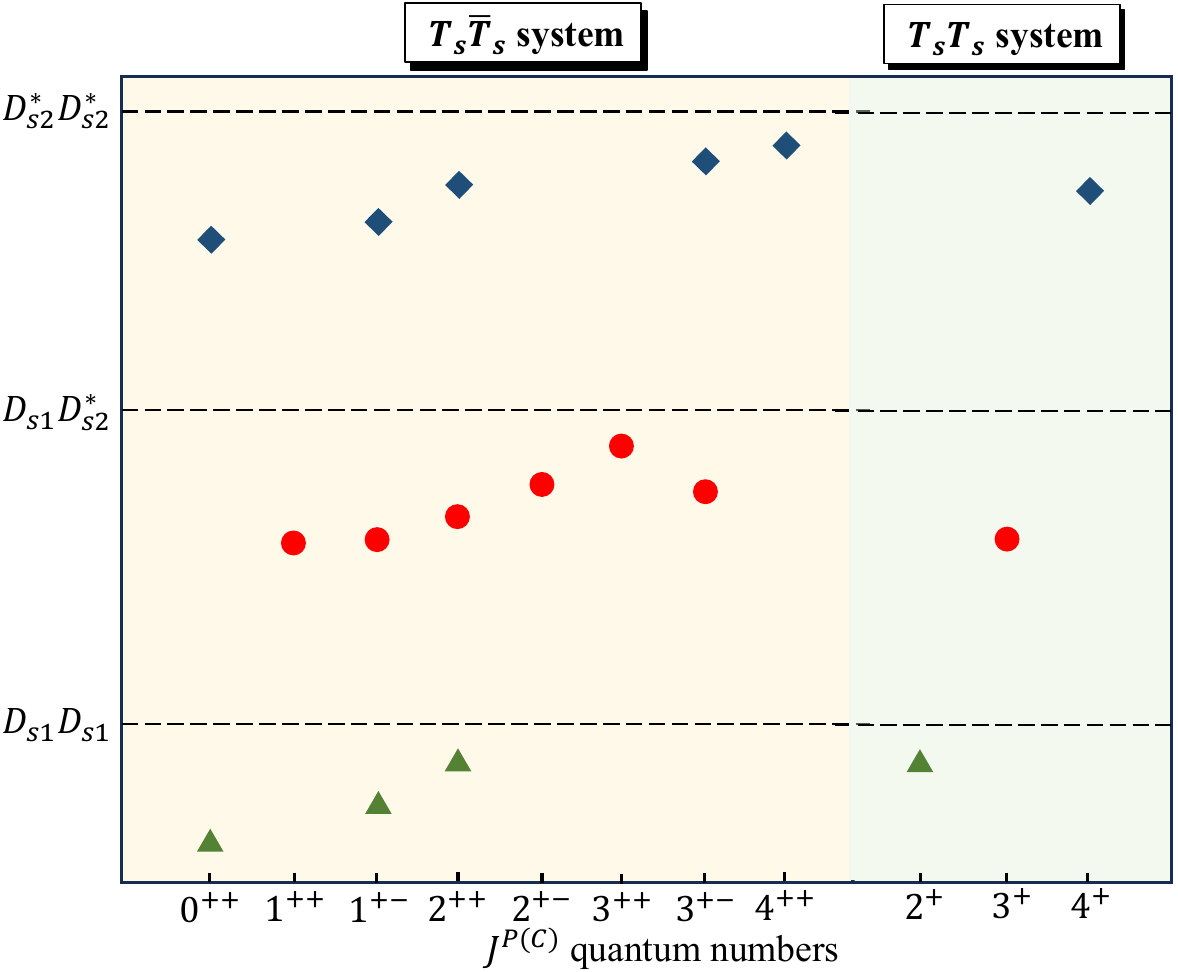}
\caption{Schematic diagram of the molecular-type characteristic spectra of the $T_s \bar T_s$ and $T_s T_s$ systems, where the horizontal dashed lines denote the threshold positions  for the corresponding channels on the vertical axis.}\label{CMS}
\end{figure}

The characteristic spectra can serve as a useful guide for the search and identification of the hadronic molecular states. A notable example is the observed hidden-charm $P_c$ states \cite{Aaij:2019vzc}, whose mass spectrum closely resembles the predicted characteristic spectrum of the $\Sigma_c \bar D^{(*)}$-type hidden-charm molecular pentaquark systems, thereby supporting their interpretation as the molecular-type exotics \cite{Li:2014gra,Karliner:2015ina,Wu:2010jy,Wang:2011rga,Yang:2011wz,Wu:2012md,Chen:2015loa}. In Fig. \ref{CMS}, we illustrate the schematic diagram of the molecular-type characteristic spectra of the $T_s \bar T_s$ and $T_s T_s$ systems. The main purpose is to indicate which hadronic molecular states are expected to exist and to reveal the overall ordering of the spectrum for the discussed systems. For the same system with different quantum numbers, we adopt identical cutoff values so that the relative binding strengths among various $J^{P(C)}$ states can be meaningfully compared. For instance, in the case of the $D_{s1} \bar D_{s1}$ system, with the same cutoff parameter, the $0^{++}$ state exhibits significantly deeper binding than the $1^{+-}$ state. Consequently, we place the $0^{++}$ state at a lower mass than the $1^{+-}$ state, reflecting this relative ordering qualitatively. Such a treatment allows us to highlight the hierarchy of the bound states without overemphasizing the accuracy of the absolute mass predictions.

It is instructive to compare our results with those for molecular tetraquarks formed from two $S$-wave charmed-strange mesons or from one $S$-wave and one $P$-wave charmed-strange mesons. In those cases, the reduced masses are smaller and the kinetic terms larger, hindering binding. Indeed, previous studies have shown that such systems require larger cutoffs to bind \cite{Wang:2021aql}. In contrast, the $T$-doublet charmed-strange mesons are considerably heavier, leading to larger reduced masses and thus smaller kinetic energies. This strongly favors the formation of the loosely bound states, as reflected in the relatively low cutoffs needed to bind our predicted candidates.

We encourage experimental searches for such predicted doubly-charm molecular tetraquarks, which can potentially be produced in the proton-proton collisions at LHCb. Drawing on the experience gained from the discovery of hidden-charm hidden-strangeness molecular tetraquark candidates in charmonium plus strange meson final states or in charmed-strange meson plus anti-charmed-strange meson channels \cite{Liu:2013waa,Hosaka:2016pey,Chen:2016qju,Richard:2016eis,Lebed:2016hpi,Liu:2019zoy,Brambilla:2019esw,Chen:2022asf,Olsen:2017bmm,Guo:2017jvc,Meng:2022ozq,Liu:2024uxn,Wang:2025sic,Wang:2025dur,Bai:2026atm}, we suggest that a similar search strategy be adopted for the $T_s\bar T_s$ systems. Moreover, recalling that the dominant decay modes of the $T$-doublet charmed-strange mesons are $D^{(*)}K$ channels \cite{ParticleDataGroup:2024cfk}, and inspired by the observation of $T_{cc}(3875)^+$ in the $DD\pi$ invariant mass spectrum \cite{LHCb:2021vvq,LHCb:2021auc}, we propose that the $T_s T_s$ systems could be searched for in final states with two charmed mesons and two kaons.

\section*{Acknowledgement}

This work is supported by the Natural Science Foundation of Gansu Province (No. 26RCKA012 and No. 25JRRA799), the National Natural Science Foundation of China under Grants No. 12335001, No. 12405097, and No. 12247101, the ‘111 Center’ under Grant No. B20063, the fundamental Research Funds for the Central Universities (lzujbky-2023-stlt01), Lanzhou City High-Level Talent Funding, and the Talent Scientific Fund of Lanzhou University.

\end{document}